\documentclass[english]{article}
\usepackage[T1]{fontenc}
\usepackage[latin9]{inputenc}
\usepackage{geometry}
\usepackage{amssymb}

\makeatletter

\newcommand{\lyxaddress}[1]{
\par {\raggedright #1
\vspace{1.4em}
\noindent\par}
}

\usepackage{babel}
\makeatother

\begin{document}

\title{Non-equilibrium statistical mechanics of classical nuclei interacting
with the quantum electron gas}

\author{Yu Wang and Lev Kantorovich}

\maketitle

\lyxaddress{Department of Physics, King's College London, The Strand, London,
WC2R 2LS, United Kingdom}

\begin{abstract}
Kinetic equations governing time evolution of positions and momenta
of atoms in extended systems are derived using quantum-classical ensembles
within the Non-Equilibrium Statistical Operator Method (NESOM). Ions
are treated classically, while their electrons quantum mechanically;
however, the statistical operator is not factorised in any way and
no simplifying assumptions are made concerning the electronic subsystem.
Using this method, we derive kinetic equations of motion for the classical
degrees of freedom (atoms) which account fully for the interaction
and energy exchange with the quantum variables (electrons). Our equations,
alongside the usual {}``Newtonian''-like terms normally associated
with the Ehrenfest dynamics, contain additional terms, proportional
to the atoms velocities, which can be associated with the {}``electronic
friction''. Possible ways of calculating the friction forces which
are shown to be given via complicated non-equilibrium correlation
functions, are discussed. In particular, we demonstrate that the correlation
functions are directly related to the thermodynamic Matsubara Green's
functions, and this relationship allows for the diagrammatic methods
to be used in treating electron-electron interaction perturbatively
when calculating the correlation functions. This work also generalises
previous attempts, mostly based on model systems, of introducing the
electronic friction into Molecular Dynamics equations of atoms. 
\end{abstract}

\section{Introduction}

Classical Molecular Dynamics (MD) simulations \cite{Allen-Tildesley-MD,Rapaport-MD}
play an important role in modern condensed matter physics \cite{LevKant-book}
giving direct access to a wide range of statistical properties of
the systems under study. In \emph{ab initio} MD simulations atoms
which are treated classically follow in time Newtonian's equations
of motion. The latter are solved numerically using atomic forces.
In \emph{ab initio} MD simulations the forces on atoms are calculated
from the first principles by considering electrons (at each atomic
configuration) entirely quantum mechanically, usually within the density
functional theory (DFT) \cite{Car-Parinello}. This approach is also
sometimes called the mean-field approximation (MFA) \cite{Prezhdo-JCP-1999}. 

Probably, the simplest quantum-mechanical justification of the MFA
\cite{Horsfield-review-2006} is based on a factorisation of the density
operator for the whole system into a \emph{product} of individual
operators for the nuclei and electrons, and then using this Ansatz
in the quantum Liouville equation with subsequent replacement of the
quantum bracket with the classical Poisson bracket for the classical
degrees of freedom. Then, a classical trajectory is introduced by
adopting a special Delta-function representation for the density operator
of the classical subsystem. The important message here is that the
ionic coordinates and momenta in the usual MD equations appear as
\emph{statistical averages} calculated at every time step. Thus, usual
MD equations constitute the dynamical equations of motion (EoM) for
averages as proposed originally a long time ago by Ehrenfest \cite{Ehrenfest-1927}.
Although more sophisticated approaches have also been developed (see
e.g. \cite{Horsfield-review-2006,Prezhdo-QHD-review-2006,Horsfield-non-adiab-2004,Darling-Holloway-review-1995}
and references therein) in which quantum nature of slow variables
(classical degrees of freedom) is taken into account to some extent,
these methods are still very complicated. At the same time, classical
consideration of nuclei can still be well justified for many problems
\cite{Head-Gordon_Tully-1995}. 

In the present paper we propose a general statistical mechanical consideration
of a system consisting of slow and fast degrees of freedom assuming
nuclei and electrons as a particular example. We derive EoM for slow
degrees of freedom (nuclei) which interact and exchange energy with
the fast degrees of freedom (electrons). Contrary to conventional
approaches (e.g. \cite{Horsfield-non-adiab-2004}) based on the Liouville
equation which possesses the time-reversal symmetry and is thus intrinsically
\emph{equilibrium} \cite{Zubarev-1}, our method is based on entirely
non-equilibrium consideration within the Non-equilibrium Statistical
Operator Method (NESOM) \cite{Zubarev-1}. The treatment of classical
and quantum degrees of freedom is done within the method of mixed
quantum-classical ensembles (MQCE) \cite{KANT_QM-NESOM_JPCM,KANT_QM-Class_ensembles}
in which the Liouville operator is represented as a sum of the classical
Poisson bracket and a quantum-mechanical commutator acting on the
statistical operator of the whole system. The latter depends on coordinates
and momenta of classical degrees of freedom and, at the same time,
is a super-operator acting on quantum operators due to quantum degrees
of freedom. A formal derivation of this method of treating mixed quantum-classical
systems based on a group-theoretical analysis was given in Refs. \cite{Aleksandrov-quant-class-bracket-1981,Prezhdo-Kisil-quant-class-bracket-1997,Prezhdo-Pereverzev-2000}.
Note that in this approach the statistical operator is not assumed
to be in a factorised form with respect to slow and fast degrees of
freedom.

Physically, one would expect that if the {}``fast'' electrons are
in instantaneous equilibrium with the {}``slow'' nuclei, then the
former would follow the dynamics of the latter and the EoM would correspond
exactly to the Ehrenfest dynamics when the nuclei move along a single
trajectory (which depends on the initial conditions) while the electrons
are in the ground state. However, in reality the electrons are quantum
particles which impose fluctuating forces on the nuclei. Also, at
a given temperature $T$, the electrons are not isolated from the
nuclei and the heat bath surrounding the system; they should get enough
energy to occupy an ensemble of ground and excited states corrresponding
to this $T$ and will require some time to equilibrate after the nuclei
displaced from their current positions. This make us think that the
motion of the nuclei cannot be considered as following a single trajectory;
instead, one can only consider the motion of the nuclei statistically,
{}``on average''. Moreover, the EoM for the nuclei average momenta
would also deviate from the Ehrenfest dynamics: firstly, the average
forces acting on the nuclei are expected to contain friction-like
terms reflecting the possibility of the energy exchange, and, secondly,
there should be a {}``conservative'' force acting on the nuclei
due to electrons occupying an ensemble of states. In this paper we
develop a general formalism that leads to this kind of description.

We show that the EoM for nuclei corresponds to the Ehrenfest dynamics
with additional terms. The latter are related to rather complicated
non-equilibrium correlation functions, and we provide a way of deriving
these terms systematically. In the first order approximation, our
additional terms are shown to be exactly proportional to atoms momenta
and can thus be interpreted as friction forces. These forces have
long been known in the literature as {}``electronic friction'' (see
review \cite{Horsfield-review-2006} and references therein), but
they were either introduced semi-empirically \cite{Finnis-Agnew-Foreman-1991},
as Langevin forces \cite{Head-Gordon_Tully-1995} or due to energy
losses in particular model systems \cite{Horsfield-review-2006,Hellsing-Persson-1984,Liebsch-1997,Plihal-Langreth-1998}.
In this paper we give a general derivation and justification of these
kind of terms. The method presented here is a generalisation of our
previous treatment \cite{KANT_QM-NESOM_JPCM,KANT_QM-Class_ensembles}
of a classical tip of Atomic Force Microscopy interacting with quantum
surface vibrations.

The plan of the paper is as follows. In the next Section we shall
introduce main concepts of the NESOM and MQCE to set up the necessary
definitions and notations. In Section 3 our main formalism is given
and the EoM for the nuclei are derived to the first order, and we
explain which quantities are used for building up this approximation.
We also discuss how this procedure can be extended systematically
to include terms up to any order. Although we do not consider any
specific model in which the non-equilibrium correlation functions
could be calculated due to an enormous complexity of those, a general
discussion on how this could be done, at least in principle, will
also be given. In particular, their connection to the Matsubara Green's
functions \cite{Abrokosov-Gorkov-Dzjaloshinskij,Zubarev-2} is discussed
in Appendix. Finally, in Section 4, main conclusions are drawn.

\section{NESOM and MQCE\label{sec:NESOM-and-MQCE}}

In NESOM statistical mechanics of a system is described in general
by a statistical operator $\rho(t)$ which satisfies the Liouville
equation with broken time-reversal symmetry \cite{Zubarev-1}:\begin{equation}
\frac{\partial\rho}{\partial t}+i\widehat{L}\rho=-\varepsilon\left(\rho-\rho_{rel}\right)\label{eq:Liuville-eq}\end{equation}
where $\widehat{L}$ is the Liouville operator and $\rho_{rel}(t)$
is the so-called \emph{relevant distribution} corresponding to local
equilibrium in the system. The right hand side of Eq. (\ref{eq:Liuville-eq})
serves to break the time-reversal symmetry inherent to the usual Liouville
equation, $\frac{\partial\rho}{\partial t}+i\widehat{L}\rho=0$, in
which this term is missing, and this guarantees that the retarded
solution of the Liouville equation is chosen corresponding to the
physically acceptable non-equilibrium behaviour of the system. The
limit $\varepsilon\rightarrow0$ (following the thermodynamic limit)
is taken after the calculation of necessary averages with respect
to $\rho(t)$. 

The key quantity in this formalism is the relevant distribution, $\rho_{rel}(t)$,
which is constructed from a set of \emph{relevant statistical variables}
$X_{n}$, and follows from the principle of maximum of the information
entropy. The entropy is maximum subject to the so-called self-consistency
conditions stating that the statistical averages of the variables
$X_{n}$ calculated with the relevant distribution, $\left\langle X_{n}\right\rangle _{rel}^{t}=\mbox{Tr}\left(\rho_{rel}(t)X_{n}\right)$,
at time $t$ are always equal to the true statistical averages, $\left\langle X_{n}\right\rangle ^{t}=\mbox{Tr}\left(\rho(t)X_{n}\right)$,
calculated with the true statistical operator, $\rho(t)$. This is
achieved using Lagrange multipliers, which makes the relevant distribution
to depend explicitly on the true statistical averages of the relevant
variables, and thus on the true statistical operator $\rho(t)$ which
is obtained by solving the modified Liouville equation (\ref{eq:Liuville-eq}).
This makes the whole scheme highly non-linear (and thus the term {}``self-consistency
conditions''). 

When considering, within the MQCE, a mixed quantum-classical system
consisting of quantum and classical degrees of freedom, one has to
use a generalised expression for the Liouville operator which acts
on classical variable $A$ as the classical Poisson bracket \begin{equation}
\left\{ A,H\right\} =\sum_{j}\left(\frac{\partial A}{\partial Q_{j}}\frac{\partial H}{\partial P_{j}}-\frac{\partial A}{\partial P_{j}}\frac{\partial H}{\partial Q_{j}}\right)\label{eq:Poisson-bracket}\end{equation}
where $Q=\left\{ Q_{j}\right\} $ and $P=\left\{ P_{j}\right\} $
are the coordinates and momenta corresponding to the classical degrees
of freedom, while it is the quantum commutator, $\frac{1}{i\hbar}[Y,H]$,
when it acts on a quantum operator $Y$ associated with the quantum
degrees of freedom. Here $H$ is the system total Hamiltonian, depending
on both types of variables at the same time. Consequently, the true
and relevant statistical operators which may also depend on both classical
and quantum variables, act as operators on quantum states and, at
the same time, are functions of coordinate and momenta of the classical
degrees of freedom, as in classical statistical mechanics. Since the
quantum operators may not necessarily commute with the Hamiltonian,
one has to use the symmetrised Poisson bracket when constructing the
appropriate generalised Liouville operator \cite{Aleksandrov-quant-class-bracket-1981,Prezhdo-Kisil-quant-class-bracket-1997,Prezhdo-Pereverzev-2000,Prezhdo-QHD-review-2006,Prezhdo-quant-class-bracket-JCP-2006},
i.e. the Liouville operator in MQCE is the sum of the quantum and
classical counterparts: \begin{equation}
i\hat{L}\ldots=i\hat{L}_{q}\ldots+i\hat{L}_{c}\ldots=\frac{1}{i\hbar}[\ldots,H]+\frac{1}{2}\left(\left\{ \ldots,H\right\} -\left\{ H,\ldots\right\} \right)\label{eq:Liouville-operator}\end{equation}
i.e. it is formally constructed as a sum of the quantum and symmetrised
classical Poisson brackets. It is readily seen that the Liouville
operator defined in this way serves as a usual commutator when acting
on quantum operators and is the classical Poisson bracket (\ref{eq:Poisson-bracket})
when acting on classical variables. If a variable contains both classical
and quantum components, the generalised operator (\ref{eq:Liouville-operator})
is to be used. 

The statistical operator in MQCE is normalised to unity in the generalised
sense, $\mbox{Tr}\left(\rho\right)=1$, via the {}``total trace''
defined as: \begin{equation}
\mbox{Tr}\left(\ldots\right)=\int\mbox{tr}(\ldots)d\Gamma\label{eq:Trace-Def}\end{equation}
where the trace written with small letters corresponds to the usual
quantum trace taken with respect to the quantum states associated
with the quantum degrees of freedom, while integration corresponds
to all coordinates and momenta of the classical phase space $\Gamma$
as in ordinary classical statistical mechanics. Correspondingly, a
statistical average of an arbitrary observable $A$, which may depend
on classical degrees of freedom and, at the same time, is an operator
in the quantum subspace, is defined in the generalised sense as\begin{equation}
\left\langle A\right\rangle ^{t}=\mbox{Tr}\left(\rho(t)A\right)\label{eq:average-of-A}\end{equation}
One can also define the {}``flux'' $\dot{A}$ operator (time derivative
of $A$) associated with the variable $A$ in the usual way as $\dot{A}=i\hat{L}A$.
It is seen that all equations look identical to either pure classical
and quantum cases, only the actual meaning of the Liouville operator
is different.

In the following, we shall limit ourselves with the Hamiltonian of
the form \begin{equation}
H=\sum_{j}\frac{P_{j}^{2}}{2M_{j}}+U(Q)+H_{q}(p,q;Q)\label{eq:Hamiltonian}\end{equation}
which corresponds to the electron-ion system in which ions of masses
$M_{j}$ are considered classically, while electrons quantum mechanically.
The index $j$ corresponds to a classical degree of freedom (i.e.
each atom contributes three such degrees of freedom). Above, the first
term gives the kinetic energy of the classical ions, their potential
energy in an external field as well as the ion-ion interaction is
provided by the second term. The last term forms the quantum Hamiltonian
for the electrons (with coordinates $q=\left\{ \mathbf{r}_{k}\right\} $
and momenta $p$),

\begin{equation}
H_{q}(p,q;Q)=H_{e}(p,q)+\Phi(Q,q)\label{eq:Hq-eltr_Hamilt}\end{equation}
This Hamiltonian describes kinetic and interaction energies of the
electrons (the first term), while their interaction with the classical
coordinates, $Q$, is described by the second term. Note that interaction
between the two subsystems in $H$ depends only on their coordinates.
Then the following expression for the classical part of the Liouville
operator acting on the operator $A$ is obtained: \begin{equation}
i\hat{L}_{c}A=\sum_{j}\left[\frac{P_{j}}{M_{j}}\frac{\partial A}{\partial Q_{j}}+\frac{1}{2}\left(\frac{\partial A}{\partial P_{j}}\dot{P}_{j}+\dot{P}_{j}\frac{\partial A}{\partial P_{j}}\right)\right]\label{eq:iLA-classical}\end{equation}
where the variable $\dot{P}_{j}=i\hat{L}P_{j}=i\hat{L}_{c}P_{j}=-\frac{\partial H}{\partial Q_{j}}$
corresponds to the force acting on the classical coordinate $Q_{j}$
(degree of freedom $j$). 

When deriving kinetic equations (equations of motion for the statistical
averages $\left\langle A\right\rangle ^{t}$), the following identity
proves to be indispensable: \begin{equation}
\mbox{Tr}\left(\left(i\hat{L}\rho\right)A\right)=-\mbox{Tr}\left(\left(i\hat{L}A\right)\rho\right)\label{eq:useful-identity}\end{equation}
where the generalised Liouville operator of Eq. (\ref{eq:Liouville-operator})
is used. Since this identity is linear with respect to the Liouville
operator, it can be proven separately for each of the Poisson brackets.
For the quantum bracket, $i\hat{L}_{q}$, it follows trivially from
the cyclic invariance of the quantum trace \cite{Zubarev-1}. To prove
it for the classical Poisson brackets, one writes:\[
\mbox{Tr}\left(\left(i\hat{L}_{c}\rho\right)A\right)=\sum_{j}\int d\Gamma\left[\frac{P_{j}}{M_{j}}\mbox{tr}\left(\frac{\partial\rho}{\partial Q_{j}}A\right)+\frac{1}{2}\mbox{tr}\left(\frac{\partial\rho}{\partial P_{j}}\dot{P}_{j}A\right)+\frac{1}{2}\mbox{tr}\left(\dot{P}_{j}\frac{\partial\rho}{\partial P_{j}}A\right)\right]\]
Using integration by parts with respect to $Q_{j}$ for the first
term in the square brackets and the fact that the density operator
should vanish at the boundaries of the phase space, we find that we
can replace the trace $\mbox{tr}\left(\frac{\partial\rho}{\partial Q_{j}}A\right)$
with $-\mbox{tr}\left(\frac{\partial A}{\partial Q_{j}}\rho\right)$.
Similar method is applied to the other two terms: using integration
by parts with respect to $P_{j}$, cyclic invariance of the quantum
trace and the fact that ionic forces $\dot{P}_{j}$ do not depend
on the momentum $P_{j}$, we obtain the following substitutions for
the second and the third traces in the square brackets above: $\mbox{tr}\left(\frac{\partial\rho}{\partial P_{j}}\dot{P}_{j}A\right)\rightarrow-\mbox{tr}\left(\dot{P}_{j}\frac{\partial A}{\partial P_{j}}\rho\right)$
and $\mbox{tr}\left(\dot{P}_{j}\frac{\partial\rho}{\partial P_{j}}A\right)\rightarrow-\mbox{tr}\left(\frac{\partial A}{\partial P_{j}}\dot{P}_{j}\rho\right)$.
This proves Eq. (\ref{eq:useful-identity}) for any quantum-classical
operator $A$.

Finally, we prove that for any general operator $\widehat{B}(P,Q)$,
acting on quantum states and depending on classical variables as well,
the following identity is satisfied: \begin{equation}
\mbox{Tr}\left(iL_{c}\widehat{B}\right)=0\label{eq:trace-identity}\end{equation}
provided that the operator $\widehat{B}$ vanishes at the boundaries
of the classical phase space. This is proven by using an explicit
expression for the classical Liouville operator, Eq. (\ref{eq:iLA-classical}).
Indeed, consider the first part of it, containing the product of the
classical momentum $P_{j}$ and the derivative $\frac{\partial\widehat{B}}{\partial Q_{j}}$.
When taking the trace, the integration over $Q_{j}$ is performed
immediately resulting in the difference $\left.\widehat{B}(P,Q)\right|_{Q_{i}=\infty}-\left.\widehat{B}(P,Q)\right|_{Q_{i}=-\infty}$
which is zero due to our assumption concerning the operator $\widehat{B}$.
Similarly, the other two terms of the classical Liouville operator,
Eq. (\ref{eq:iLA-classical}), also result in the zero contribution
due to integration over $P_{j}$ and the fact that the force, $\dot{P}_{j}$,
does not depend on the momenta.

\section{Theory\label{sec:Theory}}

\subsection{Relevant variables and distribution \label{sub:Relevant-variables}}

As we are primarily interested in this work with the equation of motion
(EoM) for the classical variables which are much slower than their
quantum counterparts, it is reasonable to sample over the fast degrees
of freedom. In practice, this is achieved by choosing classical coordinates
and momenta $Q$ and $P$ as the appropriate relevant variables. Correspondingly,
the relevant distribution maximising the information entropy at the
given temperature $T$ and number of electrons $N$ and subject to
the self-consistency conditions,\begin{equation}
\left\langle Q_{j}\right\rangle _{rel}^{t}=\left\langle Q_{j}\right\rangle ^{t},\,\,\,\left\langle P_{j}\right\rangle _{rel}^{t}=\left\langle P_{j}\right\rangle ^{t}\label{eq:scf-conditions}\end{equation}
is (cf. \cite{Zubarev-1}):\begin{equation}
\rho_{rel}(t)=\frac{1}{Z}\exp\left\{ -\beta\left[\mathcal{H}-\sum_{j}\left(V_{j}P_{j}+\mathcal{F}_{j}Q_{j}\right)\right]\right\} \label{eq:Ro-rel}\end{equation}
where $\beta=1/k_{B}T$ is the inverse temperature, $\mathcal{H}=H-\mu N$
is the system Hamiltonian containing explicitly the chemical potential
$\mu$ of electrons, $V_{j}$ and $\mathcal{F}_{j}$ are the corresponding
Lagrange multipliers and $Z$ is the normalisation factor ensuring
that $\mbox{Tr}\left(\rho_{rel}\right)=1$. Note that the sum over
$N$ is incorporated into the definition of the {}``small trace''.
The relevant statistical operator depends on time only via the Lagrange
multipliers (see below). 

At this point it is also convenient to introduce the statistical operator,
\begin{equation}
\rho_{eq}=Z_{eq}^{-1}\exp\left(-\beta\mathcal{H}_{q}\right)\label{eq:Ro-eq}\end{equation}
for the quantum subsystem, $Z_{eq}=\mbox{tr}\left(\exp\left(-\beta\mathcal{H}_{q}\right)\right)$,
where $\mathcal{H}_{q}=H_{q}-\mu N$ is the total electronic Hamiltonian
(\ref{eq:Hq-eltr_Hamilt}). It corresponds to the quantum equilibrium
canonical statistical operator for the electrons when all classical
variables are fixed (i.e. the classical subsystem is frozen). Then,
the relevant distribution can also be written as a product \begin{equation}
\rho_{rel}=\rho_{eq}f(P,Q,t)\label{eq:Ro-rel_as_Ro-eq_and_f}\end{equation}
of a {}``classical distribution function''\begin{equation}
f(P,Q,t)=\frac{Z_{eq}}{Z}\exp\left\{ -\beta\left[\sum_{j}\left(\frac{P_{j}^{2}}{2M_{j}}-V_{j}P_{j}-\mathcal{F}_{j}Q_{j}\right)+U(Q)\right]\right\} ,\,\,\,\int f(P,Q,t)d\Gamma=1\label{eq:f_of_PQt}\end{equation}
and the canonical quantum equilibrium statistical operator, $\rho_{eq}$.
We shall also need the reduced distribution function, $f(Q,t)$, which
is obtained from the distribution function above after integrating
over the momenta:\begin{equation}
f(Q,t)=\int f(P,Q,t)dP=\frac{Z_{eq}}{Z_{Q}}e^{-\beta\left(U-\sum_{j}\mathcal{F}_{j}Q_{j}\right)},\,\,\,\int f(Q,t)dQ=1\label{eq:f_of_Qt}\end{equation}
where $Z_{Q}$ is the corresponding normalisation factor. Note that
$Z_{eq}=Z_{eq}(Q)$. The average with respect to the relevant distribution
of any classical variable (depending only on classical degrees of
freedom) is obtained as the average with respect to the distribution
function $f(P,Q,t)$; if the variable depends only on the classical
coordinates, then the relevant distribution average is expressed as
the average with respect to the $Q$-only distribution $f(Q,t)$.

The Lagrange multipliers are obtained from the self-consistency conditions
(\ref{eq:scf-conditions}). Due to explicit dependence of $\rho_{rel}$
on the ions momenta via $f(P,Q,t)$, calculation of $\left\langle P_{j}\right\rangle _{rel}^{t}$
is straightforward: it gives simply $\left\langle P_{j}\right\rangle _{rel}^{t}=M_{j}V_{j}\equiv\left\langle P_{j}\right\rangle ^{t}$,
and we obtain: \begin{equation}
V_{j}=\frac{\left\langle P_{j}\right\rangle ^{t}}{M_{j}}\label{eq:Vj}\end{equation}
Thus, $V_{j}$ has the meaning of the average velocity of the degree
of freedom $j$.

Integrating with respect to all classical momenta in $\left\langle Q_{j}\right\rangle _{rel}^{t}$,
one finds that $\left\langle Q_{j}\right\rangle _{rel}^{t}=\left\langle Q_{j}\right\rangle ^{t}$
does only depend on the Lagrange multipliers $\left\{ \mathcal{F}_{j}\right\} $.
Inversely, this means that the Lagrange multipliers $\mathcal{F}_{j}$
only depend on the average coordinates $\left\{ \left\langle Q_{j}\right\rangle ^{t}\right\} $.
To obtain an explicit expression for $\mathcal{F}_{j}$, we calculate
$\left\langle \dot{P_{j}}\right\rangle _{rel}^{t}=\left\langle -\frac{\partial H}{\partial Q_{j}}\right\rangle _{rel}^{t}$,
using Eq. (\ref{eq:Ro-rel}). To this end, we first take the quantum
trace of the identity (see, e.g. \cite{Zubarev-1})\begin{equation}
\frac{\partial}{\partial Q_{j}}e^{-\beta H}=-\beta\int_{0}^{1}e^{-x\beta H}\frac{\partial H}{\partial Q_{j}}e^{x\beta H}e^{-\beta H}dx\label{eq:different-exp}\end{equation}
giving \begin{equation}
-\frac{1}{\beta}\mbox{tr}\left(\frac{\partial}{\partial Q_{j}}e^{-\beta H}\right)=\mbox{tr}\left(\frac{\partial H}{\partial Q_{j}}e^{-\beta H}\right)\label{eq:tr-identity}\end{equation}
Therefore, using this and the integration by parts with respect to
$Q_{j}$ in the $\left\langle -\frac{\partial H}{\partial Q_{j}}\right\rangle _{rel}^{t}$,
one obtains: \begin{equation}
\left\langle \dot{P_{j}}\right\rangle _{rel}^{t}=-\mathcal{F}_{j}\label{eq:Fj}\end{equation}
Thus, the second Lagrange multiplier, $\mathcal{F}_{j}$, has the
meaning of the minus average force acting on the classical degree
of freedom $j$. Using the explicit expression for the Hamiltonian,
Eq. (\ref{eq:Hamiltonian}), the ionic force $\dot{P}_{j}$ can be
broken down into conservative, \begin{equation}
F_{j}^{c}=-\frac{\partial U}{\partial Q_{j}}+\left\langle X_{j}\right\rangle _{eq}\label{eq:Fj-conservative}\end{equation}
and the stochastic, $\Delta X_{j}=X_{j}-\left\langle X_{j}\right\rangle _{eq}$,
parts, $\dot{P}_{j}=F_{j}^{c}+\Delta X_{j}$, where $X_{j}=-\frac{\partial\Phi}{\partial Q_{j}}$
is the instantaneous force acting on the degree of freedom $j$ due
to the electronic (i.e. quantum) subsystem. Therefore, using Eq. (\ref{eq:Ro-rel_as_Ro-eq_and_f}),
one obtains that \begin{equation}
\mathcal{F}_{j}=-\int f(P,Q,t)\left\langle \dot{P_{j}}\right\rangle _{eq}d\Gamma=-\int f(P,Q,t)F_{j}^{c}d\Gamma=-\left\langle F_{j}^{c}\right\rangle _{rel}^{t}\label{eq:Fj-via-Fj^c}\end{equation}
which demonstrates that the Lagrange multiplier $\mathcal{F}_{j}$
corresponds to the average of the conservative force. 

By definition, $\left\langle \Delta X_{j}\right\rangle _{eq}=0$.
However, using Eq. (\ref{eq:Ro-rel_as_Ro-eq_and_f}), one can also
check that the average of the stochastic part of the force with respect
to the relevant distribution is also zero:

\begin{equation}
\left\langle \Delta X_{j}\right\rangle _{rel}^{t}=0\label{eq:dX=0}\end{equation}

The relevant distribution has a number of properties which are proven
to be useful in our forthcoming analysis. Firstly, since $\rho_{rel}$
is equal to a product of a part, depending only on classical variables,
and the quantum operator, $e^{-\beta H_{q}}$, it commutes with the
Hamiltonian, $H$, Eq. (\ref{eq:Hamiltonian}). Therefore,\begin{equation}
i\hat{L}_{q}\rho_{rel}=0\,\,\,\mbox{and}\,\,\, e^{i\hat{L}_{q}t}\rho_{rel}(t^{\prime})=\rho_{rel}(t^{\prime})\label{eq:L_q-on-Ro_rel}\end{equation}

Next, consider the relevant distribution average $\left\langle \dot{P}_{j}\psi(Q)\right\rangle _{rel}^{t}$,
where $\psi(Q)$ is some function of the classical coordinates. Using
the explicit expression for $\rho_{rel}$, Eq. (\ref{eq:Ro-rel}),
trace identity (\ref{eq:tr-identity}) and integration by parts, one
obtains: \[
\left\langle \dot{P}_{j}\psi(Q)\right\rangle _{rel}^{t}=-\frac{1}{\beta}\left\langle \frac{\partial\psi}{\partial Q_{j}}\right\rangle _{rel}^{t}+\left\langle F_{j}^{c}\right\rangle _{rel}^{t}\left\langle \psi(Q)\right\rangle _{rel}^{t}\]
Note that $\dot{P}_{j}$ inside the angle brackets in the left hand
side of this formula can be replaced with $F_{j}^{c}$ since $\dot{P}_{j}=F_{j}^{c}+\Delta X_{j}$,
and the quantum equilibrium average of the stochastic force is equal
to zero: \[
\left\langle \Delta X_{j}\psi\right\rangle _{rel}^{t}=\int\psi(Q)f(P,Q,t)\mbox{tr}\left(\rho_{eq}\Delta X_{j}\right)d\Gamma\sim\left\langle \Delta X_{j}\right\rangle _{eq}=0\]
due to Eq. (\ref{eq:Ro-rel_as_Ro-eq_and_f}). Therefore, by taking
$\psi=Q_{i}$ and $\psi=F_{i}^{c}$, we obtain the following two useful
relationships:\begin{equation}
\left\langle F_{j}^{c}Q_{i}\right\rangle _{rel}^{t}-\left\langle F_{j}^{c}\right\rangle _{rel}^{t}\left\langle Q_{i}\right\rangle ^{t}=-\frac{1}{\beta}\delta_{ij}\label{eq:Fjc-Qi-identity}\end{equation}
\begin{equation}
\left\langle F_{j}^{c}F_{i}^{c}\right\rangle _{rel}^{t}-\left\langle F_{j}^{c}\right\rangle _{rel}^{t}\left\langle F_{i}^{c}\right\rangle _{rel}^{t}=-\frac{1}{\beta}\left\langle \frac{\partial F_{i}^{c}}{\partial Q_{j}}\right\rangle _{rel}^{t}\label{eq:Fjc-Fic-identity}\end{equation}
Since the left hand side in the second identity is symmetrical with
respect to indices $i$ and $j$, we also have the symmetry relation
for the relevant average of the derivative of the conservative force:
$\left\langle \frac{\partial F_{j}^{c}}{\partial Q_{i}}\right\rangle _{rel}^{t}=\left\langle \frac{\partial F_{i}^{c}}{\partial Q_{j}}\right\rangle _{rel}^{t}$. 

Another useful expression is obtained by differentiating both sides
of \[
\left\langle Q_{i}\right\rangle ^{t}=\left\langle Q_{i}\right\rangle _{rel}^{t}=\int\mbox{tr}\left(Q_{i}\rho_{rel}\right)d\Gamma=\int Q_{i}f(P,Q,t)d\Gamma\]
with respect to $\left\langle Q_{j}\right\rangle ^{t}$. Recalling
that only the Lagrange multipliers $\left\{ \mathcal{F}_{j}\right\} $
in $f(P,Q,t)$ (both in the exponential and in the $Z$) depend explicitly
on $\left\langle Q_{i}\right\rangle ^{t}$, we obtain: \begin{equation}
\frac{1}{\beta}\delta_{ij}=\sum_{j}\frac{\partial\mathcal{F}_{i}}{\partial\left\langle Q_{j}\right\rangle ^{t}}\mathcal{L}_{ij}\label{eq:formula1}\end{equation}
where $\mathcal{L}=\left\Vert \left\langle Q_{i}Q_{j}\right\rangle _{rel}^{t}-\left\langle Q_{i}\right\rangle ^{t}\left\langle Q_{j}\right\rangle ^{t}\right\Vert $
is a symmetric matrix. It follows from this that the derivative $\frac{\partial\mathcal{F}_{i}}{\partial\left\langle Q_{j}\right\rangle ^{t}}=\frac{\partial\mathcal{F}_{j}}{\partial\left\langle Q_{i}\right\rangle ^{t}}$
is also symmetric. Because of Eq. (\ref{eq:Fj-via-Fj^c}), the derivative
of $\left\langle F_{j}^{c}\right\rangle _{rel}^{t}$ with respect
to $\left\langle Q_{i}\right\rangle ^{t}$ is also symmetric:\[
\frac{\partial\left\langle F_{i}^{c}\right\rangle _{rel}^{t}}{\partial\left\langle Q_{j}\right\rangle ^{t}}=\frac{\partial\left\langle F_{j}^{c}\right\rangle _{rel}^{t}}{\partial\left\langle Q_{i}\right\rangle ^{t}}\]

To obtain a perturbative solution of the Liouville equation in Section
\ref{sub:Perturbative-solution-1}, some other expressions involving
the relevant distribution are needed. By differentiating $\rho_{rel}$
of Eq. (\ref{eq:Ro-rel}) with respect to the classical coordinates
and momenta, one gets: \begin{equation}
\frac{\partial\rho_{rel}(t)}{\partial P_{j}}=\frac{\beta}{M_{j}}\left(\left\langle P_{j}\right\rangle ^{t}-P_{j}\right)\rho_{rel}\label{eq:dRo-over-dPj}\end{equation}
\begin{equation}
\frac{\partial\rho_{rel}(t)}{\partial Q_{j}}=\beta\left[-\left\langle \dot{P}_{j}\right\rangle _{rel}^{t}+F_{j}^{c}+\int_{0}^{\beta}\widetilde{\Delta X}_{j}(i\lambda\hbar)d\lambda\right]\rho_{rel}\label{eq:dRo-over-dQj}\end{equation}
where we have used Eqs. (\ref{eq:Vj}), (\ref{eq:different-exp})
and (\ref{eq:Fj-via-Fj^c}), and $\widetilde{\Delta X}_{j}(x)=e^{ix\mathcal{H}_{q}/\hbar}\Delta X_{j}e^{-ix\mathcal{H}_{q}/\hbar}$
is the stochastic force in the Heisenberg representation. 

Finally, one can also calculate derivatives of $\rho_{rel}$ with
respect to the Lagrange multipliers, remembering that the {}``partition
function'' $Z$ also depends on them:\begin{equation}
\frac{\partial\rho_{rel}(t)}{\partial V_{j}}=\beta\left(P_{j}-\left\langle P_{j}\right\rangle ^{t}\right)\rho_{rel}(t)\label{eq:dRo-over-dVj}\end{equation}
\begin{equation}
\frac{\partial\rho_{rel}(t)}{\partial\mathcal{F}_{j}}=\beta\left(Q_{j}-\left\langle Q_{j}\right\rangle ^{t}\right)\rho_{rel}(t)\label{eq:dRo-over-dFj}\end{equation}

\subsection{Equations of motion for ions\label{sub:Equations-of-motion}}

In order to derive EoM for ions, we should calculate the time derivatives
of the exact averages $\left\langle P_{j}\right\rangle ^{t}$ and
$\left\langle Q_{j}\right\rangle ^{t}$. Using the Liouville equation
with broken time-reversal symmetry, Eq. (\ref{eq:Liuville-eq}), and
identity (\ref{eq:useful-identity}), one obtains:\begin{equation}
\frac{d}{dt}\left\langle Q_{n}\right\rangle ^{t}=\mbox{Tr}\left(\frac{\partial\rho(t)}{\partial t}Q_{n}\right)=\mbox{Tr}\left(\rho\left(i\hat{L}Q_{n}\right)\right)=\frac{\left\langle P_{n}\right\rangle ^{t}}{M_{n}}\label{eq:Qj-EOM}\end{equation}
where we have also used the fact that $i\hat{L}Q_{n}=i\hat{L}_{c}Q_{n}=P_{n}/M_{n}$.
Note that the term in the right hand side of the Liouville equation
(\ref{eq:Liuville-eq}) does not contribute due to the self-consistency
conditions (\ref{eq:scf-conditions}). Similarly, using Eq. (\ref{eq:useful-identity}),
we get:\begin{equation}
\frac{d}{dt}\left\langle P_{n}\right\rangle ^{t}=\mbox{Tr}\left(\rho\left(i\hat{L}P_{n}\right)\right)=\left\langle \dot{P}_{n}\right\rangle ^{t}=\left\langle \dot{P}_{n}\right\rangle _{rel}^{t}+\mbox{Tr}\left(\dot{P}_{n}\Delta\rho\right)\label{eq:Pj-EOM}\end{equation}
where $\Delta\rho=\rho-\rho_{rel}$ is the difference between the
exact and the relevant statistical operators. The obtained EoMs are
similar to Newtonian ones of the ordinary Molecular Dynamics \cite{Allen-Tildesley-MD,Rapaport-MD}
since the right hand side of Eq. (\ref{eq:Pj-EOM}) corresponds to
the actual force $\left\langle \dot{P}_{n}\right\rangle ^{t}$ acting
on degree of freedom $j$. However, this force depends, in a rather
non-trivial way, on the time evolution of the exact statistical operator
$\rho(t)$ which satisfies the Liouville equation (\ref{eq:Liuville-eq}).
Note that $\rho(t)$ is the statistical operator for the whole system,
comprising both nuclei and electrons, and no attempt has been made
to factorise $\rho$ in any way here.

The ion force $\left\langle \dot{P}_{n}\right\rangle _{rel}^{t}$
calculated with respect to the relevant distribution depends explicitly
on time via the Lagrange multipliers; the latter are some functions
of the exact expectation values $\left\langle P_{j}\right\rangle ^{t}$
and $\left\langle Q_{j}\right\rangle ^{t}$ as it has been discussed
above. Similarly, the trace in the right hand side of Eq. (\ref{eq:Pj-EOM})
would depend on the Lagrange multipliers and thus on the same expectation
values. One can see that the kinetic equations written above correspond
to some non-linear differential equations for the observables $\left\langle P_{j}\right\rangle ^{t}$
and $\left\langle Q_{j}\right\rangle ^{t}$. To obtain these equations
in the explicit form, we have to obtain an explicit expression for
$\Delta\rho(t)$ by solving the Liouville equation (\ref{eq:Liuville-eq}).
This will be done in the next subsection using a kind of a perturbation
theory (cf. Refs. \cite{KANT_QM-Class_ensembles,KANT_QM-NESOM_JPCM})
in which the square root of the relative mass of the electron and
nucleus, $\sqrt{m/M}$, is used as a small parameter.

\subsection{Perturbative solution of the Liouville equation: the first order\label{sub:Perturbative-solution-1}}

Formally, the exact solution of the Liouville equation (\ref{eq:Liuville-eq})
with respect to the $\Delta\rho(t)$ can be written as \cite{Zubarev-1}\begin{equation}
\Delta\rho(t)=-\int_{-\infty}^{0}ds\, e^{\varepsilon s}e^{is\widehat{L}}\left(\frac{\partial}{\partial r}+i\widehat{L}\right)\rho_{rel}(r)\label{eq:delta-Ro-ini}\end{equation}
where $r=t+s$. Here $\widehat{L}$ is the combined Liouville operator,
Eq. (\ref{eq:Liouville-operator}), containing both classical and
quantum parts. In order to apply the perturbation theory, we have
to calculate the quantity $\left(\frac{\partial}{\partial r}+i\widehat{L}\right)\rho_{rel}(r)$.
This calculation consists of several steps which will be outlined
below.

Firstly, $i\widehat{L}\rho_{rel}=i\widehat{L}_{c}\rho_{rel}$ due
to Eq. (\ref{eq:L_q-on-Ro_rel}). The action of the classical Liouville
operator (\ref{eq:iLA-classical}) on the relevant distribution is
obtained using Eqs. (\ref{eq:dRo-over-dPj}) and (\ref{eq:dRo-over-dQj})
and is as follows:\[
i\widehat{L}_{c}\rho_{rel}(r)=\sum_{j}\frac{\beta}{M_{j}}\left\{ \left[\left\langle P_{j}\right\rangle ^{r}F_{j}^{c}-P_{j}\left\langle F_{j}^{c}\right\rangle _{rel}^{r}+P_{j}\int_{0}^{\beta}\widetilde{\Delta X}_{j}(i\lambda\hbar)d\lambda\right]\rho_{rel}(r)\right.\]
\begin{equation}
\left.+\frac{1}{2}\left(\left\langle P_{j}\right\rangle ^{r}-P_{j}\right)\left(\Delta X_{j}\rho_{rel}(r)+\rho_{rel}(r)\Delta X_{j}\right)\right\} \label{eq:i-Lc-Ro}\end{equation}
To calculate the time derivative of the relevant distribution, we
note that it comes entirely from the Lagrange multipliers. The latter
depend on time through the observables $\left\langle P_{j}\right\rangle ^{t}$
and $\left\langle Q_{j}\right\rangle ^{t}$ which satisfy Eqs. (\ref{eq:Pj-EOM})
and (\ref{eq:Qj-EOM}), respectively. Also, we note that if $V_{j}$
depends directly only on $\left\langle P_{j}\right\rangle ^{t}$,
the other Lagrange multiplier, $\mathcal{F}_{j}$, depends on all
average coordinates $\left\{ \left\langle Q_{j}\right\rangle ^{t}\right\} $.
Therefore, \[
\frac{\partial\rho_{rel}(r)}{\partial r}=\sum_{j}\left(\frac{\partial\rho_{rel}(r)}{\partial V_{j}}\frac{\partial V_{j}}{\partial r}+\frac{\partial\rho_{rel}(r)}{\partial\mathcal{F}_{j}}\frac{\partial\mathcal{F}_{j}}{\partial r}\right)\]
\[
=\sum_{j}\frac{\beta}{M_{j}}\left(P_{j}-\left\langle P_{j}\right\rangle ^{r}\right)\left[\left\langle \dot{P}_{j}\right\rangle _{rel}^{r}+\mbox{Tr}\left(\dot{P}_{j}\Delta\rho(r)\right)\right]\rho_{rel}(r)\]
\begin{equation}
-\sum_{ij}\frac{\beta}{M_{i}}\frac{\partial\left\langle F_{j}^{c}\right\rangle _{rel}^{r}}{\partial\left\langle Q_{i}\right\rangle ^{r}}\left(Q_{j}-\left\langle Q_{j}\right\rangle ^{r}\right)\left\langle P_{i}\right\rangle ^{r}\rho_{rel}(r)\label{eq:dRo-over-t}\end{equation}
where the use have been made of Eqs. (\ref{eq:dRo-over-dVj})-(\ref{eq:Pj-EOM})
as well. 

Thus, the required quantity $\left(\frac{\partial}{\partial r}+i\widehat{L}\right)\rho_{rel}(r)$
is available now as a sum of two expressions, (\ref{eq:i-Lc-Ro})
and (\ref{eq:dRo-over-t}), given above. These are to be acted with
the exponential Liouville operator, $e^{is\widehat{L}}$ (see Eq.
(\ref{eq:delta-Ro-ini})). In turn, the Liouville operator consists
of the quantum, $i\widehat{L}_{q}$, and classical, $i\widehat{L}_{c}$,
parts. Similarly to the argument of Refs. \cite{KANT_QM-NESOM_JPCM,KANT_QM-Class_ensembles},
we argue that the classical part of $i\widehat{L}$ can be considered
as being much smaller than its quantum counterpart. Indeed, on average,
one can assume that the classical momentum $P_{j}$ is of the order
of $M^{1/2}$, where $M$ is a characteristic mass of the ions. Then,
according to Eq. (\ref{eq:iLA-classical}), $i\widehat{L}_{c}\sim M^{-1/2}$
since the forces $\dot{P}_{j}$ depend only on ionic positions, not
on their masses. Hence, the exponential operator, $e^{is\widehat{L}}=e^{s\left[i\widehat{L}_{q}+i\widehat{L}_{c}\right]}$,
can be expanded in a power series with respect to the {}``small''
Liouville operator $i\widehat{L}_{c}$. This can be done systematically
(and expressed via the time-ordered exponential operator, see, e.g.,
Chapter 6.1.1 in \cite{Zubarev-2}); here we shall only need the first
two terms:\begin{equation}
e^{is\widehat{L}}=\left[1+\int_{0}^{1}e^{ixs\widehat{L}_{q}}\left(is\widehat{L}_{c}\right)e^{-ixs\widehat{L}_{q}}dx+\ldots\right]e^{is\widehat{L}_{q}}\label{eq:exp-expansion}\end{equation}
In this Section we shall limit ourself with the very first term in
this expansion, i.e. we replace the combined Liouville operator in
Eq. (\ref{eq:delta-Ro-ini}) with its quantum part. Then, since the
action of the quantum exponential Liouville operator on any quantum
operator $A$ is simply equal to its Hermitian conjugate, $e^{is\widehat{L}_{q}}A=\widetilde{A}(s)$,
we obtain in this order of the perturbation theory, using Eqs. (\ref{eq:i-Lc-Ro})
and (\ref{eq:dRo-over-t}): \[
\Delta\rho(t)=-\int_{-\infty}^{0}ds\, e^{\varepsilon s}\left\{ \sum_{j}\frac{\beta}{M_{j}}\left[\left\langle P_{j}\right\rangle ^{r}\left(F_{j}^{c}-\left\langle F_{j}^{c}\right\rangle _{rel}^{r}\right)+\left(P_{j}-\left\langle P_{j}\right\rangle ^{r}\right)\mbox{Tr}\left(\dot{P}_{j}\Delta\rho(r)\right)\right.\right.\]
\[
\left.+P_{j}\int_{0}^{\beta}\widetilde{\Delta X}_{j}(i\lambda\hbar+s)d\lambda\right]\rho_{rel}(r)-\sum_{j}\frac{\beta}{2M_{j}}\left(P_{j}-\left\langle P_{j}\right\rangle ^{r}\right)\left(\widetilde{\Delta X}_{j}(s)\rho_{rel}(r)+\rho_{rel}(r)\widetilde{\Delta X}_{j}(s)\right)\]
\begin{equation}
\left.-\sum_{ij}\frac{\beta}{M_{i}}\left\langle P_{i}\right\rangle ^{r}\frac{\partial\left\langle F_{j}^{c}\right\rangle _{rel}^{r}}{\partial\left\langle Q_{i}\right\rangle ^{r}}\left(Q_{j}-\left\langle Q_{j}\right\rangle ^{r}\right)\rho_{rel}(r)\right\} \label{eq:delta-Ro-1st-order}\end{equation}
It is seen that this is actually an integral equation for $\Delta\rho$
since it is present inside the integral (and the trace) as well. Therefore,
it can be solved by iterations. To do this, we must analyse every
term in the above expression with respect to the small parameter of
our perturbation theory. It follows then that each term, apart from
the term with the trace, is of the order of $P/M\sim M^{-1/2}$ (because
of Eq. (\ref{eq:formula1}), the derivative $\frac{\partial\left\langle F_{j}^{c}\right\rangle _{rel}^{r}}{\partial\left\langle Q_{i}\right\rangle ^{r}}$
should be considered as of the zero order with respect to $M$). Thus,
in order to obtain $\Delta\rho(t)$ in the first order, one can simply
drop the trace term in the right hand side. We then immediately see
that this particular approximation for $\Delta\rho$, which will be
referred to in the following as $\Delta\rho_{1}$, does not break
down the normalisation of the density operator since, as it can easily
be checked by direct calculation, $\mbox{Tr}\left(\Delta\rho_{1}\right)=0$. 

Hence, dropping the trace term above, substituting the resulting expression
for $\Delta\rho_{1}$ into the kinetic equation (\ref{eq:Pj-EOM}),
and recalling that the force $\dot{P}_{n}=F_{n}^{c}+\Delta X_{n}$,
we get two terms: $\mbox{Tr}\left(F_{n}^{c}\Delta\rho_{1}(t)\right)=\int F_{n}^{c}\mbox{tr}\left(\Delta\rho_{1}\right)d\Gamma$
and $\int\mbox{tr}\left(\Delta\rho_{1}(t)\Delta X_{n}\right)d\Gamma$.
The quantum trace of $\Delta\rho_{1}$, needed for the first term,
is obtained directly from Eq. (\ref{eq:delta-Ro-1st-order}) by using
the following identities: \[
\mbox{tr}\left(\rho_{rel}(r)\right)=f(P,Q,r)\]
 \[
\mbox{tr}\left(\rho_{rel}(r)\widetilde{\Delta X}_{k}(s)\right)=f(P,Q,r)\,\mbox{tr}\left(\rho_{eq}\Delta X_{k}\right)=0\]
where the cyclic invariance of the trace was used to obtain the second
identity. Thus, we obtain:\[
\mbox{tr}\left(\Delta\rho_{1}(t)\right)=-\int_{-\infty}^{0}ds\, e^{\varepsilon s}\sum_{j}\frac{\beta}{M_{j}}\left\{ \left\langle P_{j}\right\rangle ^{r}\left(F_{j}^{c}-\left\langle F_{j}^{c}\right\rangle _{rel}^{r}\right)-\sum_{i}\left\langle P_{j}\right\rangle ^{r}\frac{\partial\left\langle F_{j}^{c}\right\rangle _{rel}^{r}}{\partial\left\langle Q_{i}\right\rangle ^{r}}\left(Q_{i}-\left\langle Q_{i}\right\rangle ^{r}\right)\right\} f(P,Q,r)\]
and, therefore,\[
\mbox{Tr}\left(F_{n}^{c}\Delta\rho_{1}(t)\right)=-\int_{-\infty}^{0}ds\, e^{\varepsilon s}\sum_{j}\frac{\beta}{M_{j}}\left\langle P_{j}\right\rangle ^{r}\left\{ \left(\left\langle F_{n}^{c}F_{j}^{c}\right\rangle _{rel}^{r}-\left\langle F_{n}^{c}\right\rangle _{rel}^{r}\left\langle F_{j}^{c}\right\rangle _{rel}^{r}\right)\right.\]
 \[
\left.-\sum_{i}\frac{\partial\left\langle F_{j}^{c}\right\rangle _{rel}^{r}}{\partial\left\langle Q_{i}\right\rangle ^{r}}\left(\left\langle F_{n}^{c}Q_{i}\right\rangle _{rel}^{r}-\left\langle F_{n}^{c}\right\rangle _{rel}^{r}\left\langle Q_{i}\right\rangle ^{r}\right)\right\} \]
This latter expression can be greatly simplified by virtue of Eqs.
(\ref{eq:Fjc-Qi-identity}) and (\ref{eq:Fjc-Fic-identity}):\begin{equation}
\mbox{Tr}\left(F_{n}^{c}\Delta\rho_{1}(t)\right)=\int_{-\infty}^{0}ds\, e^{\varepsilon s}\sum_{j}\frac{1}{M_{j}}\left[\left\langle \frac{\partial F_{j}^{c}}{\partial Q_{n}}\right\rangle _{rel}^{r}-\frac{\partial\left\langle F_{j}^{c}\right\rangle _{rel}^{r}}{\partial\left\langle Q_{n}\right\rangle ^{r}}\right]\left\langle P_{j}\right\rangle ^{r}\label{eq:trace-1}\end{equation}
Note that both derivatives inside the square brackets are symmetric
with respect to the permutation of their indices.

We shall now turn to the second term, $\mbox{Tr}\left(\Delta\rho_{1}\Delta X_{n}\right)$,
arising in the right hand side of the kinetic equation, and substitute
$\Delta\rho_{1}$ there. Noting that the trace $\mbox{Tr}\left(F_{j}^{c}\Delta X_{n}\rho_{rel}\right)$
is equal to zero due to $\mbox{tr}\left(\Delta X_{n}\rho_{eq}\right)=0$,
and the fact that \begin{equation}
\int P_{j}\rho_{rel}(r)dP_{j}=\left\langle P_{j}\right\rangle ^{r}\int\rho_{rel}(r)dP_{j}\label{eq:Pj-Ro-integral-identity}\end{equation}
which follows from the explicit dependence of $\rho_{rel}$ on the
classical momenta and Eq. (\ref{eq:Vj}), we obtain: \begin{equation}
\mbox{Tr}\left(\Delta\rho_{1}\Delta X_{n}\right)=-\int_{-\infty}^{0}ds\, e^{\varepsilon s}\sum_{j}\frac{\beta}{M_{j}}\left\langle P_{j}\right\rangle ^{r}\left(X_{n},X_{j}(s)\right)_{rel}^{r}\label{eq:trace-2}\end{equation}
where we have introduced the non-equilibrium correlation function
of the fluctuation of the ionic force (cf. \cite{KANT_QM-Class_ensembles,KANT_QM-NESOM_JPCM}):\begin{equation}
\left(X_{n},X_{k}(s)\right)_{rel}^{r}=\int_{0}^{\beta}\left\langle \Delta X_{n}\widetilde{\Delta X}_{k}(i\hbar\lambda+s)\right\rangle _{rel}^{r}d\lambda=\int dQ\, f(Q,r)\,\int_{0}^{\beta}\left\langle \Delta X_{n}\widetilde{\Delta X}_{k}(i\hbar\lambda+s)\right\rangle _{eq}d\lambda\label{eq:correl-function}\end{equation}
The last passage in the above formula is due to the fact that the
relevant distribution average under the $\lambda-$integral depends
only on the $Q$ variables since the $P$ integration can be performed
directly. Combining Eqs. (\ref{eq:trace-1}) and (\ref{eq:trace-2}),
we finally obtain the kinetic equation for the ionic momenta in the
following form:\begin{equation}
\frac{d\left\langle P_{n}\right\rangle ^{t}}{dt}=\left\langle F_{n}^{c}\right\rangle _{rel}^{t}+\sum_{j}\int_{-\infty}^{0}ds\, e^{\varepsilon s}\left\{ \left[\left\langle \frac{\partial F_{j}^{c}}{\partial Q_{n}}\right\rangle _{rel}^{r}-\frac{\partial\left\langle F_{j}^{c}\right\rangle _{rel}^{r}}{\partial\left\langle Q_{n}\right\rangle ^{r}}\right]-\beta\left(X_{n},X_{j}(s)\right)_{rel}^{r}\right\} \frac{\left\langle P_{j}\right\rangle ^{r}}{M_{j}}\label{eq:EOM-exact-1st-order}\end{equation}
This is the desired equation of motion for ions. In the right hand
side it contains the total force acting on ion $n$ due to other ions.
Their interaction and energy exchange with all the electrons are also
completely accounted for. We also note that we have not made any assumptions
as to whether electronic subsystem is in its ground state, it is in
general a weighted sum of the ground and excited electronic states
(see also below). In other words, this description goes \emph{beyond}
the adiabatic approximation. 

The first term in the right hand side of Eq. (\ref{eq:EOM-exact-1st-order})
gives the conservative force corresponding to ionic positions at time
$t$:\begin{equation}
\left\langle F_{n}^{c}\right\rangle _{rel}^{t}=\int F_{n}^{c}(Q)f(Q,t)dQ\label{eq:conserv-force-1}\end{equation}
The second term in the right hand side of Eq. (\ref{eq:EOM-exact-1st-order})
gives a correction arising due to fluctuation of the ionic force.
Similarly to the friction force acting on a Brownian particle immersed
in a liquid, this force appears to be linear with the ions momenta.
Indeed, the correlation function and derivatives inside the square
brackets do not depend on the expectation values $\left\langle P_{j}\right\rangle ^{t}$
since the $P$ integration in those terms can be performed explicitly
and all the Lagrange multipliers $\left\{ V_{j}\right\} $ disappear
exactly. Moreover, the derived friction is non-Markovian, i.e. includes
memory effects.

Thus, we conclude, the rigorous non-equilibrium statistical mechanical
treatment of a system composed of ions and electrons results in Newton-like
equations of motion for average ionic momenta that additionally contain
friction forces due to energy exchange with the electronic subsystem
maintained at the given temperature $T$. 

The obtained equations are very complicated because of the relevant
distribution used in the right hand side which depends on the observables
$\left\langle P_{j}\right\rangle ^{t}$ and $\left\langle Q_{j}\right\rangle ^{t}$
in a rather complicated way. In the next subsection a reasonable approximation
will be offered which results in a significant simplification of these
equations.

\subsection{Saddle-point approximation\label{sub:Saddle-point-approximation}}

Let us consider a relevant distribution average of some function,
$\zeta(Q)$, depending only on classical coordinates: \[
\left\langle \zeta(Q)\right\rangle _{rel}^{t}=\int\zeta(Q)f(Q,t)dQ\]
where the $Q$-distribution is given explicitly by Eq. (\ref{eq:f_of_Qt}).
Let $\left\{ \psi_{m}(q,Q);\, m=0,1,2,\ldots\right\} $ is the complete
set of electronic wavefunctions, depending parametrically on the positions
of ions $Q$. The wavefunctions $\psi_{m}$ are the eigenvectors of
the electronic Hamiltonian, i.e. $\mathcal{H}_{q}\psi_{m}=\left(\varepsilon_{m}-\mu\right)\psi_{m}$.
Then, it is easily seen that the {}``partition function'' of the
$Q-$distribution, $f(Q,t)$, can be written as a sum:\begin{equation}
Z_{Q}=\int e^{-\beta\Re(Q)}\left(1+\sum_{m=1}^{\infty}e^{-\beta\Delta\varepsilon_{m}(Q)}\right)dQ\label{eq:Z_Q-as-sum}\end{equation}
where \begin{equation}
\Re(Q)=\varepsilon_{0}(Q)-\mu+U(Q)+\sum_{j}\left\langle F_{j}^{c}\right\rangle _{rel}^{t}Q_{j}\label{eq:R(Q)}\end{equation}
and $\Delta\varepsilon_{m}(Q)=\varepsilon_{m}(Q)-\varepsilon_{0}(Q)$
are exact electronic excitation energies for the given geometry of
the nuclei, $Q$. We assume hereafter that the ground state is non-degenerate
for \emph{any} geometry, and thus all the excitation energies are
strictly positive. Moreover, we assume that for any geometry $Q$
there is a gap between the ground and the first excited states, and
the ground state energy in the external field, $\varepsilon_{0}(Q)+U(Q)$,
has a minimum at some geometry $Q_{0}$. Of course, the minimum will
be affected by the last term in Eq. (\ref{eq:R(Q)}), however, we
assume that this term does not change significantly the potential
energy surface of the ground state. Therefore, the function $\Re(Q)$
will still have a minimum at some geometry $Q^{t}$ (the subscript
reflects the fact that, because of the last term in Eq. (\ref{eq:R(Q)}),
the minimum geometry $Q^{t}$ will depend on time), and thus can be
expanded in a series with respect to the difference $Q-Q^{t}$:\[
\Re(Q)=\Re(Q^{t})+\frac{1}{2}\sum_{ij}\left(\frac{\partial^{2}\Re}{\partial Q_{i}\partial Q_{j}}\right)_{Q^{t}}\left(Q_{i}-Q_{i}^{t}\right)\left(Q_{j}-Q_{j}^{t}\right)+\ldots\]
where the matrix $\left\Vert \left(\frac{\partial^{2}\Re}{\partial Q_{i}\partial Q_{j}}\right)_{Q^{t}}\right\Vert $
of second derivatives is positively defined (since $Q^{t}$ is the
minimum). Hence, the function $e^{-\beta\Re(Q)}$ will be highly peaked
around $Q^{t}$, whereas the function in the round brackets in Eq.
(\ref{eq:Z_Q-as-sum}) can be assumed to be a rather slowly changing
with $Q$ and can thus be taken away from the $Q$-integral with all
ionic positions calculated at $Q\equiv Q^{t}$. A simple calculation
in the spirit of the well-known saddle-point approximation will then
show that the distribution function $f(Q,t)$ effectively serves as
a Delta function $\delta\left(Q-Q^{t}\right)$ giving for the average
of any slowly changing function of ionic positions, $\zeta(Q)$, the
following simple result: $\left\langle \zeta(Q)\right\rangle _{rel}^{t}\simeq\zeta(Q^{t})$.
Further, if we consider specifically $\zeta(Q)\equiv Q_{j}$, then
we obtain that $\left\langle Q_{j}\right\rangle ^{t}=\left\langle Q_{j}\right\rangle _{rel}^{t}$
should be replaced with $Q_{j}^{t}$. Thus, we conclude that in a
consistent application of the saddle point approximation, one replaces
the averages $\left\langle \zeta(Q)\right\rangle _{rel}^{t}$, calculated
with respect to the $Q-$distribution, with the corresponding functions
$\zeta$ calculated at exact average ionic positions at time $t$,
namely: \begin{equation}
\left\langle \zeta(Q)\right\rangle _{rel}^{t}\simeq\zeta\left(\left\langle Q\right\rangle ^{t}\right)\label{eq:genearl-saddle-point}\end{equation}

In particular, this result can be applied to the non-equilibrium correlation
function of Eq. (\ref{eq:correl-function}) resulting in a much simpler
expression for it: \begin{equation}
\left(X_{n},X_{k}(s)\right)_{rel}^{t}\simeq\left[\int_{0}^{\beta}\left\langle \Delta X_{n}\widetilde{\Delta X}_{k}(i\hbar\lambda+s)\right\rangle _{eq}d\lambda\right]_{{Q=\left\langle Q\right\rangle }^{t}}=\left(X_{n},X_{k}(s)\right)_{eq}^{\left\langle Q\right\rangle ^{t}}\label{eq:correl-function-eq}\end{equation}
Thus, the correlation function depends directly on a single time $s$;
however, the statistical average is to be calculated over the equilibrium
distribution $\rho_{eq}$ corresponding to ions fixed in their exact
positions $\left\langle Q\right\rangle ^{t}$ at another time $t$. 

Using the same approach, one can also verify that the term in the
square brackets in the right hand side of the equation of motion (\ref{eq:EOM-exact-1st-order})
can be dropped. Indeed, since the conservative forces $F_{j}^{c}$
depend entirely on ionic positions, we can write: \[
\left\langle \frac{\partial F_{j}^{c}}{\partial Q_{n}}\right\rangle _{rel}^{r}\simeq\left(\frac{\partial F_{j}^{c}}{\partial Q_{n}}\right)_{Q=\left\langle Q\right\rangle ^{r}}=\frac{\partial F_{j}^{c}\left(\left\langle Q\right\rangle ^{r}\right)}{\partial\left\langle Q_{n}\right\rangle ^{r}}\]
 and, at the same time, \[
\frac{\partial\left\langle F_{j}^{c}\right\rangle _{rel}^{r}}{\partial\left\langle Q_{n}\right\rangle ^{r}}\simeq\frac{\partial F_{j}^{c}\left(\left\langle Q\right\rangle ^{r}\right)}{\partial\left\langle Q_{n}\right\rangle ^{r}}\]
Therefore, within the saddle-point approximation, the difference of
derivatives in the square brackets in the equation of motion (\ref{eq:EOM-exact-1st-order})
is equal to zero. Following the same arguments and replacing the conservative
force $\left\langle F_{n}^{c}\right\rangle _{rel}^{t}$ in the equation
of motion with $F_{n}^{c}\left(\left\langle Q\right\rangle ^{t}\right)$,
we obtain:

\begin{equation}
\frac{d\left\langle P_{n}\right\rangle ^{t}}{dt}=F_{n}^{c}\left(\left\langle Q\right\rangle ^{t}\right)-\sum_{j}\frac{\beta}{M_{j}}\int_{-\infty}^{0}ds\, e^{\varepsilon s}\left(X_{n},X_{j}(s)\right)_{eq}^{\left\langle Q\right\rangle ^{t}}\left\langle P_{j}\right\rangle ^{r}\label{eq:EOM-saddle-p}\end{equation}
which is the final result. 

We see that, if not for the friction term, the equations of motion
would have corresponded exactly to the Newton's equations of motion
for ions: in the left hand side we have the time derivative of ion
$n$ momentum, while in the right hand side - the total statistically
averaged force acting on this ion at the given temperature:\[
F_{n}^{c}\left(\left\langle Q\right\rangle ^{t}\right)=\left\langle -\frac{\partial H}{\partial Q_{n}}\right\rangle _{eq}^{Q=\left\langle Q\right\rangle ^{t}}=-\frac{\partial U}{\partial Q_{n}}+\left[Z_{eq}^{-1}\sum_{m}e^{-\beta\left(\varepsilon_{m}-\mu\right)}\left\langle \psi_{m}\right|-\frac{\partial H_{q}}{\partial Q_{n}}\left|\psi_{m}\right\rangle \right]_{Q=\left\langle Q\right\rangle ^{t}}\]
 \begin{equation}
=-\frac{\partial U}{\partial Q_{n}}+\left[Z_{eq}^{-1}\sum_{m}e^{-\beta\left(\varepsilon_{m}-\mu\right)}\left(-\frac{\partial\varepsilon_{m}}{\partial Q_{n}}\right)\right]_{Q=\left\langle Q\right\rangle ^{t}}\label{eq:final-conserv-force}\end{equation}
where $Z_{eq}=\sum_{m}\exp\left(-\beta\left(\varepsilon_{m}-\mu\right)\right)$.
Here, $-\frac{\partial\varepsilon_{m}}{\partial Q_{n}}$ is the force
acting on ion $n$ (due to electrons and nuclei) when the electronic
subsystem is in electronic state $m$ (i.e. on the adiabatic potential
energy surface, $\varepsilon_{m}(Q)$, corresponding to electronic
state $m$). 

Note that atomic positions, $\left\langle Q\right\rangle ^{t}$, correspond
exactly to the averaged ionic momenta, $\left\langle P\right\rangle ^{t}$,
see Eq. (\ref{eq:Qj-EOM}). The obtained equations of motion are more
general than those of ordinary Molecular Dynamics. Indeed, in standard
MD the forces do not depend on temperature and are calculated as in
Eq. (\ref{eq:final-conserv-force}) from the conservative part taking
only the electronic ground state into account, \[
F_{n}^{c}\simeq-\frac{\partial U}{\partial Q_{n}}+\left\langle \psi_{0}\right|-\frac{\partial H_{q}}{\partial Q_{n}}\left|\psi_{0}\right\rangle =-\frac{\partial\left(U+\varepsilon_{0}\right)}{\partial Q_{n}}\]
(the Car-Parinello ground state \emph{ab initio} MD simulations \cite{Car-Parinello}
being an obvious example). Therefore our equations may serve as a
justification of MD simulations which go beyond the Born-Oppenheimer
approximation (see, e.g. \cite{Head-Gordon_Tully-1995}), where, when
calculating the force acting on an ion, it is assumed explicitly that
the electronic subsystem may occupy both ground and excited electronic
states. In fact, in complete agreement with the principles of quantum
statistical mechanics, we show that there is a certain probability
for the electronic subsystem to occupy every state at the same time,
which, one must admit, is somewhat different from most of the non-adiabatic
computational techniques \cite{Horsfield-review-2006} in which it
is usually assumed that only one state can be occupied at every single
time step.

We also observe that a consistent non-equilibrium treatment results
in additional terms in the equations of motion which are proportional
to the ions momenta and thus have the meaning of friction, related
to the energy exchange between the ions and the electrons; the latter
serving as a {}``thermostat'' held at a given temperature. Thus,
our rigorous treatment justifies the usage of {}``electronic friction''
terms in MD simulations \cite{Head-Gordon_Tully-1995} and explains
their physical origin.

\subsection{Non-equilibrium correlation functions \label{sub:Non-equilibrium-correlation-function}}

Since the operator of the atomic force, $X_{i}=-\frac{\partial\Phi}{\partial Q_{i}}=\sum_{k=1}^{N}x_{i}(\mathbf{r}_{k})$,
is a derivative of the electron-phonon interaction energy, $\Phi(q,Q)=\sum_{k=1}^{N}\phi(\mathbf{r}_{k},Q)$,
it is a one-particle operator, \[
X_{i}=\sum_{ab}X_{ab}^{i}c_{a}^{\dagger}c_{b}\]
(where $c_{a}^{\dagger}$ and $c_{b}$ are creation and annihilation
operators in some basis set of spin-orbitals) and $X_{ab}^{i}=\left\langle a\right|x_{i}(\mathbf{r})\left|b\right\rangle $.
Here $x_{i}(\mathbf{r})=-\frac{\partial\phi(\mathbf{r},Q)}{\partial Q_{i}}$
is the force on atom $i$ due to a single electron at $\mathbf{r}$,
$\phi(\mathbf{r},Q)$ being the interaction energy of this electron
with all nuclei. Hence in general, the correlation function (\ref{eq:correl-function-eq})
is a two-particle equilibrium statistical average, containing four
$c$-operators, and thus cannot be calculated exactly in the general
case.

The calculation is straightforward if the Hamiltonian $H_{q}=\sum_{ab}h_{ab}c_{a}^{\dagger}c_{b}$
is a one-particle operator (e.g. in the Hartree-Fock approximation).
Indeed, in this case one can diagonalise the Hamiltonian,\[
H_{q}=\sum_{\sigma}\xi_{\sigma}d_{\sigma}^{\dagger}d_{\sigma}\]
where \begin{equation}
d_{\sigma}=\sum_{a}e_{\sigma a}^{*}c_{a},\,\,\, d_{\sigma}^{\dagger}=\sum_{a}e_{\sigma a}c_{a}^{\dagger}\label{eq:correspond-d-c-opers}\end{equation}
and $\xi_{\sigma}$ and $e_{\sigma}=\left\Vert e_{\sigma a}\right\Vert $
are the eigenvalues and the eigenvectors of the matrix $h=\left\Vert h_{ab}\right\Vert $,
and therefore express the operators $X_{i}$ and $X_{j}$ via the
operators $d_{\sigma}^{\dagger}$ and $d_{\sigma}$, e.g.\begin{equation}
X_{i}=\sum_{\sigma\sigma^{\prime}}x_{\sigma\sigma^{\prime}}^{(i)}d_{\sigma}^{\dagger}d_{\sigma^{\prime}},\,\,\, x_{\sigma\sigma^{\prime}}^{(i)}=\sum_{ab}X_{ab}^{i}e_{\sigma a}^{*}e_{\sigma^{\prime}b}\label{eq:binary-operators}\end{equation}
Since the calculation of the operators $\widetilde{d}_{\sigma}^{\dagger}(t)$
and $\widetilde{d}_{\sigma}(t)$ in the Heisenberg representation
is simple, we obtain:\begin{equation}
\left(X_{i},X_{j}(t)\right)_{eq}=\sum_{\sigma\sigma^{\prime}}x_{\sigma\sigma^{\prime}}^{(i)}x_{\sigma^{\prime}\sigma}^{(j)}n_{\sigma}\left(1-n_{\sigma^{\prime}}\right)\chi\left(\xi_{\sigma^{\prime}}-\xi_{\sigma}\right)e^{i\left(\xi_{\sigma^{\prime}}-\xi_{\sigma}\right)t/\hbar}\label{eq:CF-in-TB}\end{equation}
where $n_{\sigma}=\left(e^{\beta(\xi_{\sigma}-\mu)}+1\right)^{-1}$
and $\chi(E)=E^{-1}\left(1-e^{-\beta E}\right)$ if $E\neq0$ and
$\chi(0)=\beta$. Note that the superscript $\left\langle Q\right\rangle ^{t}$
to the correlation function has been omitted to simplify the notations. 

When the electron-electron interaction is accounted for explicitly,
one has to develop more powerful methods. First we note that the correlation
function, Eq. (\ref{eq:correl-function-eq}), may be considered as
a particular case of a more general correlation function defined for
any two quantum operators $A$ and $B$ as follows: \begin{equation}
\left(A(t_{1}),B(t_{2})\right)_{eq}=\int_{0}^{\beta}d\lambda\,\mbox{tr}\left[\widetilde{A}(t_{1})\widetilde{B}(t_{2}+i\lambda\hbar)\rho_{eq}\right]\equiv\int_{0}^{\beta}d\lambda\left\langle \widetilde{A}(t_{1})\widetilde{B}(t_{2}+i\lambda\hbar)\right\rangle _{eq}\label{eq:corr-fun-symmetrical}\end{equation}
This correlation function obeys some simple symmetry properties: \begin{equation}
\left(A(t_{1}),B(t_{2})\right)_{eq}=\left(A,B(t_{2}-t_{1})\right)_{eq}=\left(A(t_{2}),B(t_{1})\right)_{eq}\label{eq:corr-fun-symm-prop}\end{equation}
which follow from the cyclic invariance of the trace (change of variables
$\lambda\rightarrow\lambda_{1}=\beta-\lambda$ is also necessary to
obtain the last equality). 

An explicit expression for the correlation function can be obtained
using the complete set, $\left\{ \psi_{n}\right\} $, of the eigenvectors
of the electronic Hamiltonian (i.e. $H_{q}\psi_{n}=\varepsilon_{n}\psi_{n}$):
\begin{equation}
\left(A,B(t)\right)_{eq}=\sum_{nm}\rho_{n}\chi(E_{m}-E_{n})\left\langle \psi_{n}\right|A\left|\psi_{m}\right\rangle \left\langle \psi_{m}\right|B\left|\psi_{n}\right\rangle e^{i\left(\varepsilon_{m}-\varepsilon_{m}\right)t/\hbar}\label{eq:CF-expanded}\end{equation}
where $\rho_{n}=Z_{eq}^{-1}e^{-\beta\left(\varepsilon_{n}-\mu N\right)}$.
It is also convenient to introduce the spectral function (matrix)
as the Fourier transform of the correlation function:\begin{equation}
\left(A,B(t)\right)_{eq}=\frac{1}{2\pi}\int_{-\infty}^{\infty}e^{i\omega t}J_{AB}(\omega)d\omega\label{eq:CF-FT}\end{equation}
where \begin{equation}
J_{AB}(\omega)=2\pi\hbar\sum_{nm}\rho_{n}\chi(E_{m}-E_{n})\left\langle \psi_{n}\right|A\left|\psi_{m}\right\rangle \left\langle \psi_{m}\right|B\left|\psi_{n}\right\rangle \delta\left(\varepsilon_{m}-\varepsilon_{n}-\hbar\omega\right)\label{eq:CF-spectral-F}\end{equation}
which satisfy the following symmetry properties: \begin{equation}
J_{AB}(\omega)=J_{BA}(-\omega)=J_{BA}(\omega)^{*}\label{eq:CF-SF-property}\end{equation}

Unfortunately, it appears impossible to develop systematically a perturbation
theory for the direct calculation of the correlation function (\ref{eq:corr-fun-symmetrical}).
It has been found, however, that such a perturbation theory exists
for the equilibrium statistical average inside the $\lambda-$integral,
i.e for the $\left\langle \widetilde{A}(t_{1})\widetilde{B}(t_{2}+i\lambda\hbar)\right\rangle _{eq}$.
Once the statistical average is calculated, the correlation function
 follows immediately. The method is based on a relationship between
the correlation function above and the Matsubara Green's functions
(e.g. \cite{Abrokosov-Gorkov-Dzjaloshinskij,Zubarev-2}); this is
revealed in Appendix. Therefore, one can use the perturbation expansion
(and thus the powerful diagrammatic techniques) to account for the
electron-electron interactions in calculating the correlation function
appearing in Eq. (\ref{eq:correl-function-eq}).

\subsection{Perturbative solution of the Liouville equation: second order\label{sub:Perturbative-2nd-order}}

The formulae developed in the previous Sections correspond to the
first order approximation as we kept only terms of the order of $M^{-1/2}$.
This treatment can be systematically extended to higher orders. Unfortunately,
this results in very cumbersome expressions even in the second order.
Therefore, in this Section we shall simply outline the main idea of
how the extension to higher orders can be done. 

We start from Eq. (\ref{eq:delta-Ro-ini}) and replace the exponential
operator there with its expansion given by Eq. (\ref{eq:exp-expansion}).
After that, we recall that the operator $\left(\frac{\partial}{\partial r}+i\widehat{L}\right)\rho_{rel}(r)$
contains the trace of $\Delta\rho$ as well (see Eq. (\ref{eq:dRo-over-t}))
and thus this dependence must also be included in developing the perturbative
expansion. For instance, one gets for the next order term: \[
\Delta\rho_{2}(t)=-\int_{-\infty}^{0}ds\, e^{\varepsilon s}\sum_{j}\frac{\beta}{M_{j}}\left(P_{j}-\left\langle P_{j}\right\rangle ^{r}\right)\rho_{rel}(r)\,\mbox{Tr}\left(\dot{P}_{j}\Delta\rho_{1}(r)\right)\]
\begin{equation}
-\int_{-\infty}^{0}ds\, e^{\varepsilon s}\int_{0}^{1}dx\, e^{ixs\widehat{L}_{q}}\left(is\widehat{L}_{c}\right)e^{-ixs\widehat{L}_{q}}e^{-is\widehat{L}_{q}}\widehat{A}\label{eq:delta-Ro-2}\end{equation}
where $\widehat{A}$ is obtained by removing the trace term in the
operator $\left(\frac{\partial}{\partial r}+i\widehat{L}\right)\rho_{rel}(r)$
(see Eqs. (\ref{eq:i-Lc-Ro}) and (\ref{eq:dRo-over-t})) and $\Delta\rho_{1}$
is the first order term given by Eq. (\ref{eq:delta-Ro-1st-order}). 

First of all, it can easily be shown that $\mbox{Tr}\left(\Delta\rho_{2}\right)=0$,
i.e. this correction is consistent as well with the correct normalisation
of the statistical operator. Indeed, the trace of the first term in
Eq. (\ref{eq:delta-Ro-2}) vanishes due to integration over $P_{j}$
in the classical part of the trace and Eq. (\ref{eq:Pj-Ro-integral-identity}).
The second term also does not contribute to the trace of $\Delta\rho_{2}$
due to the cyclic invariance of the trace and the operator identity
(\ref{eq:trace-identity}) (the operator $\widehat{A}$ is proportional
to the relevant distribution and thus vanishes at the boundaries of
the classical phase space). 

It can also be shown that any order correction to the statistical
operator, $\Delta\rho_{n}$, has zero trace and hence does not break
down the normalisation of the statistical operator.

Using the explicit expression for $\Delta\rho_{2}$ given above, one
can calculate its contribution to the force, $\mbox{Tr}\left(\dot{P}_{n}\Delta\rho_{2}(t)\right)$,
in the right-hand side of the equations of motion (\ref{eq:Pj-EOM}).
The contribution is very cumbersome (and is only due to the second
term in Eq. (\ref{eq:delta-Ro-2})) and won't be reproduced here.
We only note that it is proportional to the square of the atomic momenta,
i.e. it contains terms proportional to $\left\langle P_{j}\right\rangle ^{r}\left\langle P_{k}\right\rangle ^{r}$.
Higher order terms contain more products of the atomic momenta. Also,
much more complicated correlation functions appear as the kernels
of the time integrals in the contribution to the force.

\section{Conclusions}

In this paper we have considered, using intrinsically non-equilibrium
statistical mechanical theory, a system of fast (electrons) and slow
(atoms) particles which interact with each other and may interchange
their energy. The system is enclosed in a thermal bath kept at a constant
temperature. In our treatment, the combined quantum-classical consideration
was used: the slow degrees of freedom (atomic coordinates) were treated
classically, while the fast variables (electrons) quantum mechanically.
No assumption was made concerning the structure of the statistical
operator; in particular, it is not in any way factorised. In addition,
the electronic subsystem was treated exactly with complete inclusion
of electron-electron interaction. 

We show, by assuming that the classical degrees of freedom are much
{}``heavier'' than the quantum ones, that equations of motion for
the former (i.e for atoms) contain the conservative and friction forces.
The conservative forces are statistically averaged over the electronic
states. The friction force, which is strictly proportional to the
atoms momenta, is expressed via the correlation function of the fluctuating
force with which electrons act on the atoms (i.e. due to the fluctuation
of the electron-phonon interaction). The correlation function can
be expressed via two-electron Matsubara Green's function and thus
calculated using the well-developed perturabtion diagrammatic techniques. 

The theory presented here gives a solid foundation for a number of
intuitive theories based on MD simulations which go beyond the Born-Oppenheimer
approximation (see, e.g. \cite{Head-Gordon_Tully-1995,Horsfield-review-2006}).
We also justify the usage of {}``electronic friction'' terms in
MD simulations \cite{Head-Gordon_Tully-1995} and explain their physical
origin. 

In our method it was assumed that on average the electronic subsystem
is in thermodynamic equilibrium. For instance, the theory developed
here is applicable to ordinary Molecular Dynamics when atoms move
along classical trajectories. However, the theory can also be applied
in other cases in which the electronic subsystem is in a steady state,
i.e. which, on average (more precisely, over the timescale associated
with atomic motion), is not time dependent (is stationary), for instance,
stationary electronic or heat conductance in an atomic wire \cite{Horsfield-non-adiab-2004,Horsfield-review-2006}. 

A number of avenues exist in developing our theory further. Firstly,
as has been mentioned above, we have assumed in our treatment that
electrons quickly reach thermodynamic equilibrium during the motion
of atoms. In some cases this assumption will not be valid, e.g. when
considering non-elastic effects during a nonstationary conductance
in a system. In these and similar cases one has to include additional
relevant variables into the consideration, e.g. time dependent electronic
density matrix as in the kinetic theories \cite{Zubarev-1}. This
approach would result in an additional kinetic equation for the density
to be solved simultaneously with the atomic motion considered here.
Secondly, another possible extention is concerned with considering
nuclei quantum mechanically as well. This, however, is much more difficult
and may be done by e.g. starting from the method developed in Ref.
\cite{Horsfield-non-adiab-2004}.

\section*{Acknowledgemnents}

Yu Wang would like to acknowledge the financial support from the China
Scholarship Council, The K. C. Wong Education Foundation (Hong Kong),
The Henry Lester Trust Limited and The Leche Trust, which made this
work possible. We would also like to thank Tchavdar Todorov, Andrew
Horsefield and Mike Finnis for stimulating discussions and useful
suggestions.

\section*{Appendix}

In this Appendix we shall relate the correlation function (\ref{eq:correl-function-eq})
with the thermodynamic Matsubara Green's function (e.g. \cite{Abrokosov-Gorkov-Dzjaloshinskij,Zubarev-2}).
We shall start by defining a {}``complex time'' Green's function
for two arbitrary operators $A$ and $B$ as follows:\begin{equation}
G_{AB}(x_{1},x_{2})=-\mbox{tr}\left[\rho_{eq}\widehat{T}_{x}\left(\overline{A}(x_{1})\overline{B}(x_{2})\right)\right]\label{eq:GF-Def}\end{equation}
where the quantum operators $A$ and $B$ appearing in the Green's
function are written in the special representation defined with respect
to a single real parameter $x$. This specific representation which
will be designated in what follows with the bar over the operator
symbol, is given by\begin{equation}
\overline{A}(x)=e^{i\tau(x)\mathcal{H}_{q}/\hbar}Ae^{-i\tau(x)\mathcal{H}_{q}/\hbar}\label{eq:new-Heisenberg-repr}\end{equation}
where the complex {}``time'' $\tau(x)=\zeta x$ is introduced which
contains the complex prefactor $\zeta=-\left(t+i\lambda\hbar\right)/\lambda$
and changes linearly with $x$, so that $\tau(0)=0$ and $\tau(-\lambda)=t+i\lambda\hbar$.
The operator $\widehat{T}_{x}$ performs chronological ordering of
the operators $\overline{A}(x_{1})$ and $\overline{B}(x_{2})$, so
that $x$ increases from right to left:\begin{equation}
\widehat{T}_{\lambda}\left(\overline{A}(x_{1})\overline{B}(x_{2})\right)=\theta\left(x_{1}-x_{2}\right)\overline{A}(x_{1})\overline{B}(x_{2})+\eta\theta\left(x_{2}-x_{1}\right)\overline{B}(x_{2})\overline{A}(x_{1})\label{eq:T-operator-Def}\end{equation}
where $\eta=\pm1$ corresponds to the sign acquired when changing
the order of the operators $\overline{A}(x_{1})$ and $\overline{B}(x_{2})$;
$\theta(x)$ is the Heviside step function. 

Using the cyclic invariance of the trace, it is easy to see that the
Green's function (\ref{eq:GF-Def}) actualy depends only on the difference
$x=x_{1}-x_{2}$:

\[
G_{AB}(x_{1},x_{2})\equiv G_{AB}(x)=-\theta\left(x\right)\left\langle A\overline{B}(-x)\right\rangle _{eq}-\eta\theta\left(-x\right)\left\langle \overline{B}(-x)A\right\rangle _{eq}\]
The correlation function we would like to calculate, $\left\langle A\widetilde{B}(t+i\lambda\hbar)\right\rangle _{eq}$
with $0<\lambda<\beta$, is equal directly to the minus Green's function
$G_{AB}(0,-\lambda)\equiv G_{AB}(\lambda)$. Moreover, using eigenvectors
and eigenvalues of the Hamiltonian, $\mathcal{H}_{q}\psi_{n}=\left(\varepsilon_{n}-\mu N\right)\psi_{n}$,
one obtains:\begin{equation}
\left\langle A\overline{B}(-x)\right\rangle _{eq}=\sum_{nm}\rho_{n}\left\langle \psi_{n}\right|A\left|\psi_{m}\right\rangle \left\langle \psi_{m}\right|B\left|\psi_{n}\right\rangle e^{i\left(\varepsilon_{m}-\varepsilon_{n}\right)xt/\lambda\hbar}e^{-x\left(\epsilon_{m}-\epsilon_{n}\right)},\,\,\, x>0\label{eq:CF-pos-x}\end{equation}

\begin{equation}
\left\langle \overline{B}(-x)A\right\rangle _{eq}=\sum_{nm}\rho_{n}\left\langle \psi_{n}\right|A\left|\psi_{m}\right\rangle \left\langle \psi_{m}\right|B\left|\psi_{n}\right\rangle e^{i\left(\varepsilon_{m}-\varepsilon_{n}\right)xt/\lambda\hbar}e^{-(x+\beta)\left(\epsilon_{m}-\epsilon_{n}\right)},\,\,\, x<0\label{eq:CF-neg-x}\end{equation}
where $\rho_{n}=Z_{eq}^{-1}e^{-\beta\left(\epsilon_{n}-\mu N\right)}$.
It is easily verified (by splitting each double sum into two contributions
with positive energy differences each) that the first correlation
function, Eq. (\ref{eq:CF-pos-x}), converges for $x<\beta$ (and
$x>0$), while the second one for $x>-\beta$ (and $x<0$). Therefore,
$x$ is limited to the interval $-\beta<x<\beta$ . 

Note also that the above correlation functions can also be written
as Fourier integrals:\[
G_{AB}(x)=-\left\langle A\overline{B}(-x)\right\rangle _{eq}=\intop_{-\infty}^{\infty}\frac{d\omega}{2\pi}I_{AB}(\omega)e^{i\omega xt/\lambda}e^{-x\omega\hbar},\,\,\, x>0\]
\[
G_{AB}(x)=-\eta\left\langle \overline{B}(-x)A\right\rangle _{eq}=\eta\intop_{-\infty}^{\infty}\frac{d\omega}{2\pi}I_{AB}(\omega)e^{i\omega xt/\lambda}e^{-(x+\beta)\omega\hbar},\,\,\,\, x<0\]
where the spectral function\[
I_{AB}(\omega)=-2\pi\sum_{nm}\rho_{n}\left\langle \psi_{n}\right|A\left|\psi_{m}\right\rangle \left\langle \psi_{m}\right|B\left|\psi_{n}\right\rangle \delta\left(\frac{\epsilon_{m}-\epsilon_{n}}{\hbar}-\omega\right)\]

So far, the Green's function introduced above in Eq. (\ref{eq:GF-Def})
has been shown to possess properties very similar or even identical
to those of the Matsubara Green's function \cite{Abrokosov-Gorkov-Dzjaloshinskij,Zubarev-2}.
To strengthen this analogy, one can also expand the Green's function
into a Fourier series in the interval $-\beta<x<\beta$ or notice
that the two Green's functions, for $-\beta<x<0$ and $0<x<\beta$,
make a jump at $x=0$:\[
\lim_{\delta\rightarrow+0}\left[G_{AB}(\delta)-G_{AB}(-\delta)\right]=\intop_{-\infty}^{\infty}\frac{d\omega}{2\pi}I_{AB}(\omega)\left(1-\eta e^{-\beta\hbar\omega}\right)\]
However, this analogy is not complete; for instance, one can see from
the integral representations of the Green's function given above that
$G_{AB}(x)$ for $x<0$ is not related to the $G_{AB}(x+\beta)$ although
they both share the same spectral function $I_{AB}(\omega)$. 

At this point we are quite prepared to find the relationship between
the {}``complex time'' and the Matsubara Green's functions. To this
end, we shall first introduce the appropriate {}``interaction representation''
for the operators: \[
\overline{A}_{I}(x)=e^{i\tau(x)\mathcal{H}_{0}/\hbar}Ae^{-i\tau(x)\mathcal{H}_{0}/\hbar}\]
where $\mathcal{H}_{0}$ is the Hamiltonian of non-interacting electrons,
i.e. $\mathcal{\mathcal{H}}_{q}=\mathcal{H}_{0}+H^{\prime}$ with
$H^{\prime}$ being the electron-electron interaction. Then, the product
of two operators in the Green's function (\ref{eq:GF-Def}) can be
written as\begin{equation}
\overline{A}(x_{1})\overline{B}(x_{2})=U(0,x_{1})\overline{A}_{I}(x_{1})U(x_{1},x_{2})\overline{B}_{I}(x_{2})U(x_{2},0)\label{eq:AB-product-via-U}\end{equation}
where \begin{equation}
U(x_{1},x_{2})=e^{i\tau(x_{1})\mathcal{H}_{0}/\hbar}e^{-i\left(\tau(x_{1})-\tau(x_{2})\right)\mathcal{H}_{q}/\hbar}e^{-i\tau(x_{2})\mathcal{H}_{0}/\hbar}\label{eq:U-operator}\end{equation}
is the evolution operator in our interaction representation. The evolution
operator defined above satisfies the differential equation ($\kappa=\frac{i\zeta}{\hbar}=1-\frac{it}{\hbar\lambda}$)\[
\frac{\partial U(x_{1},x_{2})}{\partial x_{1}}=-\kappa\overline{H}_{I}^{\prime}(x_{1})U(x_{1},x_{2})\]
which can be converted into an integral equation and then solved iteratively,
giving as a solution: \begin{equation}
U(x_{1},x_{2})=\widehat{T}_{x}\exp\left(-\kappa\int_{x_{2}}^{x_{1}}\overline{H}_{I}^{\prime}(x)dx\right)\label{eq:U-operator-solved}\end{equation}
Thus, the evolution operator can be expressed via the $T_{x}$-exponent.
Note that $\kappa$ actually depends on the particular values of $t$
and $\lambda$ used to define the {}``complex times'' in the Green
's function arguments.

The fact that the evolution operator can be expressed via the chronologically
ordered exponent allows for the considerable simplification of the
chronological product of the two operators in the Green's function.
Indeed, the product in Eq. (\ref{eq:GF-Def}) can be rewritten as\[
\widehat{T}_{x}\left(\overline{A}(x_{1})\overline{B}(x_{2})\right)=\widehat{T}_{x}\left(U(0,x_{1})\overline{A}_{I}(x_{1})U(x_{1},x_{2})\overline{B}_{I}(x_{2})U(x_{2},0)\right)\]
\[
=\widehat{T}_{x}\left(U(0,x_{1})U(x_{1},x_{2})U(x_{2},0)\overline{A}_{I}(x_{1})\overline{B}_{I}(x_{2})\right)\]
\begin{equation}
=\widehat{T}_{x}\left(U(0,0)\overline{A}_{I}(x_{1})\overline{B}_{I}(x_{2})\right)=\widehat{T}_{x}\left(\overline{A}_{I}(x_{1})\overline{B}_{I}(x_{2})\right)\label{eq:chron-product-of-A-B}\end{equation}
where use has been made of the obvious properties of the evolution
operator: $U(x,x)=1$ and $U(x_{1},x_{2})U(x_{2},x_{3})=U(x_{1},x_{3})$.
Thus, we see that the electron-electron interaction can be actually
eliminated entirely from the chronological product of the operators
in the Green's function.

Although the method developed below is valid for any operators $A$
and $B$, we shall consider the particular case needed here when the
operators are of the binary form given by Eq. (\ref{eq:binary-operators})
(note also that $\eta=1$ in this case). To proceed, we recognise
that the creation and annihilation operators, $d_{\sigma}^{\dagger}$
and $d_{\sigma}$ (in the representation that diagonalises the Hamiltonian
$\mathcal{H}_{0}$, see Section \ref{sub:Non-equilibrium-correlation-function}),
have a very simple form both in the standard thermodynamic (imaginary
time) and our (complex time) representations. To simplify the notations,
we shall use from now on in this Appendix a wavy line above operators
for the usual thermodynamic interaction representation of the operators,
i.e. $\widetilde{C}_{I}(x)=e^{x\mathcal{H}_{0}}Ce^{-x\mathcal{H}_{0}}$.
Then, one has:\[
\overline{d}_{\sigma I}(x)=d_{\sigma}e^{-i\tau(x)\xi_{\sigma}/\hbar},\,\,\,\overline{d}_{\sigma I}^{\dagger}(x)=d_{\sigma}^{\dagger}e^{i\tau(x)\xi_{\sigma}/\hbar}\]
\[
\widetilde{d}_{\sigma I}(x)=e^{x\mathcal{H}_{0}}d_{\sigma}e^{-x\mathcal{H}_{0}}=d_{\sigma}e^{-x\xi_{\sigma}},\,\,\,\widetilde{d}_{\sigma I}^{\dagger}(x)=d_{\sigma}^{\dagger}e^{x\xi_{\sigma}}\]
Therefore, any binary operator, $A=\sum_{\sigma\sigma^{\prime}}A_{\sigma\sigma^{\prime}}d_{\sigma}^{\dagger}d_{\sigma^{\prime}}$,
when written in the {}``complex time'' interaction representation,
$\overline{A}_{I}(x)$, can easily be expressed as another operator
$A^{\prime}$ written in the ordinary thermodynamic representation,
i.e. \[
\overline{A}_{I}(x)=\sum_{\sigma\sigma^{\prime}}A_{\sigma\sigma^{\prime}}\overline{d}_{\sigma}^{\dagger}(x)\overline{d}_{\sigma^{\prime}}(x)=\sum_{\sigma\sigma^{\prime}}A_{\sigma\sigma^{\prime}}^{\prime}(x)\widetilde{d}_{\sigma}^{\dagger}(x)\widetilde{d}_{\sigma^{\prime}}(x)\equiv\widetilde{A}_{I}^{\prime}(x)\]
with the new matrix of the coefficients $\mathbf{A^{\prime}}(x)=\left\Vert e^{(\kappa-1)x\left(\xi_{\sigma}-\xi_{\sigma^{\prime}}\right)}A_{\sigma\sigma^{\prime}}\right\Vert $
which depends explicitly on $x$. This simple result allows us to
rewrite the chronological product of the operators of Eq. (\ref{eq:chron-product-of-A-B})
as \begin{equation}
\widehat{T}_{x}\left(\overline{A}(x_{1})\overline{B}(x_{2})\right)=\widehat{T}_{x}\left(\overline{A}_{I}(x_{1})\overline{B}_{I}(x_{2})\right)=\widehat{T}_{x}\left(\widetilde{A}_{I}^{\prime}(x_{1})\widetilde{B}_{I}^{\prime}(x_{2})\right)\label{eq:chron-product-final}\end{equation}
In the final expression above the product of two operators appears
exactly as in the thermodynamic (Matsubara) Green's function. To finish
the transformation, we should introduce the evolution operator in
the usual thermodynamic representation:\begin{equation}
\widetilde{U}(x_{1},x_{2})=e^{x_{1}\mathcal{H}_{0}}e^{-\left(x_{1}-x_{2}\right)\mathcal{H}_{q}}e^{-x_{2}\mathcal{H}_{0}}\equiv\widehat{T}_{x}\exp\left[-\int_{x_{2}}^{x_{1}}\widetilde{H}_{I}^{\prime}(x)dx\right]\label{eq:U-oper-therm-repr}\end{equation}
that satisfies the properties $\widetilde{U}(x_{1},x_{2})\widetilde{U}(x_{2,},x_{3})=\widetilde{U}(x_{1},x_{3})$
and $\widetilde{U}(x,x)=1$. The evolution operator enters the equilibrium
statistical operator, $\rho_{eq}$ \cite{Abrokosov-Gorkov-Dzjaloshinskij,Zubarev-2}:
\begin{equation}
\rho_{eq}=Z_{eq}^{-1}e^{-\beta\left(\mathcal{H}_{0}+H^{\prime}\right)}=\rho_{0}\frac{\widehat{T}_{x}\exp\left[-\int_{0}^{\beta}\widetilde{H}_{I}^{\prime}(x)dx\right]}{\left\langle \widehat{T}_{x}\exp\left[-\int_{0}^{\beta}\widetilde{H}_{I}^{\prime}(x)dx\right]\right\rangle ^{0}}=\frac{Z_{0}}{Z_{eq}}\rho_{0}\widetilde{U}(\beta,0)\label{eq:Ro-eq-via-T}\end{equation}
where $Z_{eq}=Z_{0}\left\langle \widetilde{U}(\beta,0)\right\rangle ^{0}$
and the brackets $\left\langle \ldots\right\rangle ^{0}=\mbox{tr}\left[\rho_{0}\ldots\right]$
correspond to the statistical average with respect to $\rho_{0}=Z_{0}^{-1}e^{-\beta\mathcal{H}_{0}}$
with $Z_{0}=\mbox{tr}\left(e^{-\beta\mathcal{H}_{0}}\right)$. 

The following steps depend on the particular values of the arguments
$x_{1}$ and $x_{2}$. However, since the Green's function depends
only on their difference, $x=x_{1}-x_{2}$, which lies between -$\beta$
and $\beta$, it is sufficient to consider only negative values of
$x_{1}$ and $x_{2}$. Thus, combining Eqs. (\ref{eq:chron-product-final})
and (\ref{eq:Ro-eq-via-T}), we can write for the product of the operators
in the Green's function (\ref{eq:GF-Def}):\[
\rho_{eq}\widehat{T}_{x}\left(\overline{A}(x_{1})\overline{B}(x_{2})\right)=\frac{Z_{0}}{Z_{eq}}\rho_{0}\widetilde{U}(\beta,0)\widehat{T}_{x}\left[\widetilde{A}_{I}^{\prime}(x_{1})\widetilde{B}_{I}^{\prime}(x_{2})\right]\]
The evolution operator $\widetilde{U}(\beta,0)$ contains the sum
of ordered products of operators $\widetilde{H}_{I}^{\prime}(x)$
whose arguments $x$ lie between zero and $\beta$, i.e. are all \emph{positive}.
Since, by our assumption, both $x_{1}$ and $x_{2}$ are \emph{negative},
the above formula can be transformed into:\begin{equation}
\frac{Z_{0}}{Z_{eq}}\rho_{0}\widehat{T}_{x}\left[\widetilde{U}(\beta,0)\widetilde{A}_{I}^{\prime}(x_{1})\widetilde{B}_{I}^{\prime}(x_{2})\right]\label{eq:product-1}\end{equation}
which results in the following final expression for the Green's function:

\begin{equation}
G_{AB}(x_{1},x_{2})=-\frac{\left\langle \widehat{T}_{x}\widetilde{U}(\beta,0)\widetilde{A}_{I}^{\prime}(x_{1})\widetilde{B}_{I}^{\prime}(x_{2})\right\rangle ^{0}}{\left\langle \widetilde{U}(\beta,0)\right\rangle ^{0}}=-\frac{\left\langle \widehat{T}_{x}\exp\left[-\int_{0}^{\beta}\widetilde{H}_{I}^{\prime}(x)dx\right]\widetilde{A}_{I}^{\prime}(x_{1})\widetilde{B}_{I}^{\prime}(x_{2})\right\rangle ^{0}}{\left\langle \widehat{T}_{x}\exp\left[-\int_{0}^{\beta}\widetilde{H}_{I}^{\prime}(x)dx\right]\right\rangle ^{0}}\label{eq:GF-connect-Matsubara}\end{equation}
which is nothing but the Matsubara Green's function, $\mathcal{G}_{A^{\prime}B^{\prime}}(x_{1},x_{2})$.
The latter is defined with respect to the operators $A^{\prime}$
and $B^{\prime}$ which are obtained from the original operators $A$
and $B$ by using the primed matrices of coefficients as explained
above. 

Thus, there is a direct connection between the Green's function (\ref{eq:GF-Def})
and the appropriate Matsubara Green's function. Since there is a well-known
diagrammatic perturbation technique developed for the latter with
the denominator cancelling out exactly as a prefactor to connected
diagrams in the nominator \cite{Abrokosov-Gorkov-Dzjaloshinskij,Zubarev-2},
\begin{equation}
G_{AB}(x_{1},x_{2})=-\left\langle \widehat{T}_{x}\exp\left[-\int_{0}^{\beta}\widetilde{H}_{I}^{\prime}(x)dx\right]\widetilde{A}_{I}^{\prime}(x_{1})\widetilde{B}_{I}^{\prime}(x_{2})\right\rangle _{c}^{0}\,\,,\label{eq:GF-connected-diagr}\end{equation}
where the subscript {}``c'' indicates explicitly that only connected
diagrams are to be retained, this method can be directly used to obtain
corrections beyond the one-electron approximation. The latter was
employed in Section \ref{sub:Non-equilibrium-correlation-function}
to derive the formula for the correlation function. In particular,
in the zero order (when the exponential operator above is replaced
by unity), the same expression is obtained for the correlation function
as in Section \ref{sub:Non-equilibrium-correlation-function}. Notice
that the direct application of Eq. (\ref{eq:GF-connected-diagr})
results in an expression containing an additional term with the product
of averages $\left\langle A\right\rangle ^{0}\left\langle B\right\rangle ^{0}$;
this term did not appear in Section \ref{sub:Non-equilibrium-correlation-function}
since the correlation funciton considered there contained already
the difference operators $\Delta A$ and $\Delta B$. \bibliographystyle{/Users/lev/PAPERS/physrev}

\begin{thebibliography}{10}

\bibitem{Allen-Tildesley-MD}
M.~P. Allen and D.~J. Tildesley,
\newblock {\em Computer simulation of liquids} (Clarendon Press, Oxford, 2002).

\bibitem{Rapaport-MD}
D.~C. Rapaport,
\newblock {\em The art of molecular dynamics simulation} (Cambridge Univ.
  press, Cambridge, 2002).

\bibitem{LevKant-book}
L.~Kantorovich,
\newblock {\em Quantum theory of the solid state: an introduction.}Fundamental
  Theories of Physics (Kluwer Academic Publishers, Dordrecht, 2004).

\bibitem{Car-Parinello}
R.~Car and M.~Parrinello,
\newblock Phys. Rev. Lett. {\bf 55}, 2471 (1985).

\bibitem{Prezhdo-JCP-1999}
O.~V. Prezhdo,
\newblock J. Chem. Phys. {\bf 111}, 8366 (1999).

\bibitem{Horsfield-review-2006}
A.~P. Horsfield {\em et~al.},
\newblock Rep. Prog. Phys. {\bf 69}, 1195 (2006).

\bibitem{Ehrenfest-1927}
P.~Ehrenfest,
\newblock Z. Phys. {\bf 45}, 455 (1927).

\bibitem{Prezhdo-QHD-review-2006}
O.~V. Prezhdo,
\newblock Theor. Chem. Accounts {\bf 116}, 206 (2006).

\bibitem{Horsfield-non-adiab-2004}
A.~P. Horsfield, D.~R. Bowler, A.~J. Fisher, T.~N. Todorov, and C.~G.
  S\'anchez,
\newblock J. Phys. Condens. Matter {\bf 16}, 8251 (2004).

\bibitem{Darling-Holloway-review-1995}
G.~R. Darling and S.~Holloway,
\newblock Rep. Prog. Phys. {\bf 58}, 1595 (1995).

\bibitem{Head-Gordon_Tully-1995}
M.~Head-Gordon and J.~C. Tully,
\newblock J. Chem. Phys. {\bf 103}, 10137 (1995).

\bibitem{Zubarev-1}
D.~Zubarev, V.~Morozov, and G.~R\"opke,
\newblock {\em Statistical mechanics of nonequilibrium processes. Vol. 1: Basic
  concepts, kinetic theory} (Akademie verlag, Berlin, 1996).

\bibitem{KANT_QM-NESOM_JPCM}
L.~N. Kantorovich,
\newblock J. Phys.: Condens. Matter {\bf 14}, 7123 (2002).

\bibitem{KANT_QM-Class_ensembles}
L.~N. Kantorovich,
\newblock Phys. Rev. Lett. {\bf 89}, 096105 (2002).

\bibitem{Aleksandrov-quant-class-bracket-1981}
I.~V. Aleksandrov,
\newblock Z. Naturforsch A {\bf 36A}, 902 (1981).

\bibitem{Prezhdo-Kisil-quant-class-bracket-1997}
O.~V. Prezhdo and V.~V. Kisil,
\newblock Phys. Rev. A {\bf 56}, 162 (1997).

\bibitem{Prezhdo-Pereverzev-2000}
O.~V. Prezhdo and Y.~V. Pereverzev,
\newblock J. Chem. Phys. {\bf 113}, 6557 (2000).

\bibitem{Finnis-Agnew-Foreman-1991}
M.~Finnis, P.~Agnew, and J.~E. Foreman,
\newblock Phys. Rev. B {\bf 44}, 567 (1991).

\bibitem{Hellsing-Persson-1984}
B.~Hellsing and M.~Persson,
\newblock Physica Scripta {\bf 29}, 360 (1984).

\bibitem{Liebsch-1997}
A.~Liebsch,
\newblock Phys. Rev. B {\bf 55}, 13263 (1997).

\bibitem{Plihal-Langreth-1998}
M.~Plihal and D.~C. Langreth,
\newblock Phys. Rev. B {\bf 58}, 2191 (1998).

\bibitem{Abrokosov-Gorkov-Dzjaloshinskij}
A.~A. Abrikosov, L.~P. Gorkov, and I.~E. Dzyaloshinski,
\newblock {\em Methods of Quantum Field Theory in Statistical Physics} (Dover,
  N. Y., 1975).

\bibitem{Zubarev-2}
D.~Zubarev, V.~Morozov, and G.~R\"opke,
\newblock {\em Statistical mechanics of nonequilibrium processes. Vol. 2:
  Relaxation and hydrodynamic processes} (Akademie verlag, Berlin, 1996).

\bibitem{Prezhdo-quant-class-bracket-JCP-2006}
O.~V. Prezhdo,
\newblock J. Chem. Phys. {\bf 124}, 201104 (2006).

\end{thebibliography}

\end{document}